\documentclass[letter]{article}
\pdfoutput=1

\usepackage{soul}
\usepackage{bm}
\usepackage{appendix}
\usepackage{multirow}
\usepackage[many]{tcolorbox}
\usepackage{empheq}
\tcbset{
  highlight math style={
    enhanced,
    colframe=red!60!black,
    colback=yellow!50,
    arc=4pt,
    boxrule=1pt,
    drop fuzzy shadow
  }
}
\usepackage[caption=false]{subfig}
\usepackage{amsmath,amsfonts,relsize}
\usepackage{graphicx}
\usepackage{dcolumn}
\usepackage[mathlines]{lineno}
\usepackage{diagbox}
\usepackage{tabularx}
\usepackage{color,xcolor}
\usepackage{comment}
\usepackage{floatrow}
\usepackage{afterpage}
\usepackage{setspace}
\usepackage{enumitem}
\usepackage{xspace}
\usepackage{slashed,physics}
\usepackage{braket}
\usepackage{float}
\usepackage{mathtools}
\usepackage{jcappub}  
\usepackage[nameinlink,capitalize]{cleveref}

\crefname{equation}{Eq.}{Eqs.}

\newcommand{\MeV}{\ensuremath{\mathrm{MeV}}}
\newcommand{\GeV}{\ensuremath{\mathrm{GeV}}}
\newcommand{\TeV}{\ensuremath{\mathrm{TeV}}}

\newcommand{\SM}{\ensuremath {\mathrm{SM}}}

\newcommand{\braketm}[2]{\mathinner{\langle #1 \mid #2 \rangle}}
\def\fd{f_{D}}
\def\deltaN{\Delta N_{D}}
\def\mpd{m_{p_{D}}}
\def\med{m_{e_{D}}}
\def\alphad{\alpha_{D}}
\def\mud{\mu_{D}}
\def\mHd{m_{H_{D}}}
\def\bd{B_{H_{D}}}

\def\rdao{r_{\rm{DAO}}}

\def\lcdm{\Lambda \text{CDM}}

\title{
Pushing the Limits of Atomic Dark Matter: First-Principles Recombination Rates and Cosmological Constraints
}

\author[a]{Jared Barron,}
\author[a]{Rouven Essig,}
\author[a,b]{Megan H. McDuffie,}
\author[b]{Jes\'{u}s P\'{e}rez-R\'{i}os,}
\author[a,b]{Gregory Suczewski}

\affiliation[a]{C.N. Yang Institute for Theoretical Physics, Stony Brook University, Stony Brook, NY 11794}
\affiliation[b]{Physics and Astronomy Department, Stony Brook University, NY 11794, USA}

\emailAdd{jared.barron@stonybrook.edu}
\emailAdd{rouven.essig@stonybrook.edu}
\emailAdd{megan.mcduffie@stonybrook.edu}
\emailAdd{jesus.perezrios@stonybrook.edu}
\emailAdd{gregory.suczewski@stonybrook.edu}

\abstract{
Minimal atomic dark matter with its distinctive cooling mechanisms offers an instructive framework for understanding the potential impact of dark matter on small-scale structure formation and early cosmology. The model consists of two fermions with opposite charges under a hidden Abelian gauge symmetry $U(1)_D$ and masses $\mpd$ and $\med$, respectively. Analogous to hydrogen in the Standard Model, these fermions interact via their own electromagnetic-like force, with a dark fine structure constant denoted by $\alphad$, and can bind into neutral atomic (and molecular) dark states. Previous work has largely focused on the benchmark scenario where the dark sector mirrors ordinary matter, with $\med$ near the electron mass, $\mpd$ near the proton mass, and $\alphad \sim 1/137$. We extend this analysis by investigating dark recombination and cooling physics across the full parameter space of masses and couplings. Combining Cosmic Microwave Background (CMB) measurements from Planck and ACT with BAO and Pantheon+ data, we place new constraints on the atomic dark matter parameter space, identifying regions where acoustic damping and recombination dynamics leave observable imprints on the CMB.
}

\begin{document}

\maketitle

\section{Introduction}
The standard cosmological model of dark energy plus cold, collisionless dark matter (called $\lcdm$) matches cosmological and astrophysical observations very well at large scales and high redshifts~\cite{Planck:2018vyg,ACT:2025fju}. However, at small scales and later times its predictions are less well-tested by data, and there are potential discrepancies with observations that allow for and motivate dark sector models that deviate from the cold, collisionless paradigm~\cite{Bullock:2017xww,Abdalla:2022yfr,Kamionkowski:2022pkx,Riess_2022,Finkelstein_2023, Harikane_2023, bouwens_evolution_2023,adame_desi_2025, w4c6-1r5j}. 

One such model is atomic dark matter (aDM), wherein two new fermionic species interact through a dark $U(1)_{D}$ gauge interaction and can form hydrogen-like neutral bound states~\cite{Kaplan:2009de,Kaplan:2011yj}. In analogy with the Standard Model (SM) sector, these two species are called a dark proton and dark electron, although there is no dark QCD-equivalent in the minimal model (which will be our focus). The interactions between dark matter and the massless dark photon can cause dark acoustic oscillations (DAOs) in the matter power spectrum at small scales~\cite{Cyr-Racine:2012tfp,Cyr-Racine:2013fsa} as well as dissipation of energy in galaxies~\cite{Mohapatra:2001sx,Fan:2013yva,Foot:2013vna,Foot:2014mia,Foot:2016wvj,Foot:2017dgx,Foot:2018qpw,Roy:2023zar,Roy:2024bcu}, affecting the shape of dark matter halos and possibly allowing fragmentation into compact objects~\cite{Ghalsasi:2017,Gurian_2022aas,Buckley:2024eoe,Buckley:2017ttd,Chang:2018bgx,Ryan:2021dis,Gurian:2021qhk,Ryan:2021tgw}. The presence of dark radiation also contributes to $\Delta N_{\mathrm{eff}}$, impacting both the Cosmic Microwave Background (CMB) damping tail and primordial elemental abundances~\cite{Bansal:2022qbi}. In order to satisfy constraints on self-interactions, a partially interacting dark sector is often considered, with only a fraction of the dark matter possessing non-gravitational self-interactions. The aDM model has been investigated for its potential to resolve the cosmological $H_{0}$ and $S_{8}$ tensions as well as galactic-scale discrepancies between the predictions of CDM and observations~\cite{Foot:2013vna,Foot:2015mqa,Foot:2016wvj,Chacko:2016kgg,Blinov:2021,Bansal:2022,Bansal:2022qbi,Hughes:2023tcn,Buen-Abad:2022kgf, Buen-Abad:2024tlb}.

The rich phenomenology of the aDM model can be constrained with many different cosmological and astrophysical observables. Studies using the CMB and large-scale structure have placed strong bounds on the model parameter space down to fractional abundances $\lesssim 5\%$~\cite{Cyr-Racine:2013fsa,Bansal:2022,Bansal:2022qbi,Hughes:2023tcn}. Recent work using measurements of the high-redshift galactic UV luminosity function as a probe of structure extended these constraints to smaller scales ($k\sim 10-100h/\mathrm{Mpc}$) for larger abundances~\cite{Barron:2025dys}. The galactic-scale phenomenology due to dissipation has been studied with analytical methods and with simulations~\cite{Fan:2013yva, Roy:2023zar,Gemmell:2023trd,Roy:2024bcu}, finding that even a small fraction of dark matter being strongly dissipative can significantly disrupt galactic morphology. Searches for gravitational wave signals from sub-solar mass black holes formed by collapsing aDM fragments have also been used to derive constraints~\cite{Shandera:2018xkn,Singh:2020wiq,LVK:2022ydq}.

Since the distinctive phenomenology that sets aDM apart from both CDM and other dark sector models arises from its composite nature and interactions with dark radiation, the cross sections and rates of atomic processes such as recombination, stimulated emission and absorption of dark photons, and collisional excitation are crucial ingredients for the types of analyses mentioned above. Furthermore, because the particle content and interactions of the model are analogous at the electromagnetic level to those of SM protons and SM electrons, we can make use of the fact that the rates for all of these processes have been calculated within the SM. These foundational calculations in atomic physics make use of two key simplifying assumptions, which hold to high precision in the SM: first, that the proton is much heavier than the electron, such that its dynamics can be ignored; and second, that the fine structure constant $\alphad$ is much less than one, permitting second- and higher-order terms to be dropped in the perturbative expansion of the interaction. The resulting rates typically have simple power-law dependence on the fine structure constant, electron mass, and proton mass. The equivalent rates in an atomic dark sector with different values of the gauge coupling and masses can then be obtained by simply rescaling by the appropriate powers of the ratios of each parameter with the SM values. This method has been used in previous studies of aDM~\cite{Cyr-Racine:2012tfp, Fan:2017,DarkKromeI,DarkKromeII}. However, the calculation of the SM rates is predicated on the assumptions mentioned previously and in an atomic dark sector there are no \textit{a priori} restrictions on the particle masses or couplings that guarantee these approximations hold.  For example, the dark proton and dark electron masses could be equal, realizing the `positronium-like' limit, and the dark fine structure constant need not be small. 

There is therefore an outstanding question of how far these re-scaled SM rates can be extrapolated to high dark electron--dark proton mass ratios and fine structure constants, and at what point, if at all, the scaling relations break down. This uncertainty has limited the range of parameters that have been explored in previous studies of aDM. In this work we investigate this question, focusing on the radiative processes that contribute to cosmological recombination. Explicit calculations of recombination coefficients in the SM have focused on two mass limits, hydrogen-like with $\mpd \gg \med$~\cite{Burgess:1960, Burgess1965}, and positronium-like with $\mpd=\med$~\cite{Erdas:1985}, and assumed a coupling strength much smaller than unity. The first major aim of this paper is to compute recombination coefficients and bound-bound transition rates for neutral dark hydrogen from first principles, allowing for arbitrary dark electron and dark proton masses, as well as $\alphad \lesssim 0.3$. The theoretical framework presented is correct until perturbation theory breaks down (ie. $\alphad \sim 1$); however, as $\alphad$ grows, fine structure and relativistic effects become large, and therefore we only calculate recombination coefficients and transition rates to $\alphad \lesssim 0.3$, where these effects remain subdominant. 

We compare the precision calculation to the SM scalings over a range of temperatures for dark electron--dark proton mass ratios from the hydrogen-like to positronium-like limits and $\alphad \leq 0.3$. Previous studies~\cite{Bansal:2022qbi} have constrained aDM assuming SM rates can be extrapolated up to $\alpha_D \lesssim 0.2$ and $\med/\mpd \lesssim 1/10$, without explicit verification. In this work, we find that the rescaled SM recombination coefficients as well as atomic transition rates that are relevant for case-B recombination (in which electrons initially transition to the $n=2$ state and only then to the $1s$ ground state) differ from our first-principles result below the $\mathcal{O}(10\%)$ level. As we will show, the SM scalings are worse for bound-to-bound transitions directly to the ground state ($1s$) at relatively low cosmological temperatures, where the largest deviations are $\mathcal{O}(1)$; however, these transitions do not directly impact the case-B recombination rates. 

The cosmological observables most sensitive to dark recombination due to the formation of DAOs are the CMB and large scale structure. We find that these observables are negligibly modified if the atomic rates deviate by less than $\mathcal{O}(1)$ from the SM scalings. Armed with this knowledge, we can confidently use the simple re-scaled SM rates (rather than the computationally intensive first-principles calculation) to compute recombination rates in aDM across a wider parameter space than what was studied in previous work, extending to the positronium-like limit ($\med/\mpd=1$) and $\alphad\leq$ 0.3.  This allows us to fulfill the second major aim of this paper, which is to use new measurements of CMB anisotropies to unprecedentedly large $\ell$ by the Atacama Cosmology Telescope~\cite{ACT:2025fju} to calculate up-to-date large-scale cosmological constraints on aDM across the entire parameter space where the perturbative and non-relativistic assumptions hold. 

In~\autoref{sec:aDM_model}, we introduce the aDM model and the atomic processes whose rates we compute. In~\autoref{sec:QM}, we solve the two-body Schr\"odinger equation for the system of a dark proton and dark electron, introduce the interaction potential, and calculate the bound and continuum wavefunctions. 
In \autoref{sec:extFieldsinQM}, we then use these wavefunctions to calculate the cross sections for bound-bound and continuum-bound radiative transitions. 
In~\autoref{sec:CompareSMScalings}, we compare our first-principles bound-bound and continuum-bound transition rates to those obtained using SM scalings. In~\autoref{sec:cosmology}, we present new constraints on the aDM parameter space from measurements of the CMB by the ACT and Planck telescopes, as well as from large-scale structure data. We conclude in~\autoref{sec:Conclusion}. We show an explicit calculation of the relevant overlaps in~\autoref{sec:overlaps}. We provide physical intuition for the origin of the deviations from the SM scalings as well as present improved 
scalings in~\autoref{sec:DeviationsFromSM}.
%%%%%%%%%%%%%%%%%%%%%%%%%%%%%%%%%%%%%%%%%%%%%%%%%%%%%%%%%%%%%%%%%%%%%%%%%%%%%%%%%%%%
\section{Atomic Dark Matter and its interactions}
\label{sec:aDM_model}

We consider a model containing two Dirac fermion fields that are oppositely charged under a hidden $U(1)_D$ massless gauge field with field strength tensor $A_{\mu \nu}$ and gauge coupling $e_D$. In analogy with the SM, the two fermions are called the dark electron and dark proton, although the dark proton does not possess any dark QCD-like interactions. The dark electron and dark proton masses are $\med$ and $\mpd$, respectively, and $\alphad=e^{2}_D/4\pi$ is the dark fine structure constant. An asymmetric relic abundance of dark electrons and dark protons accounts for some or all of the dark matter with fractional abundance $\fd$, with the remainder assumed to be cold and collisionless. Finally, a thermal background of dark photons exists with temperature $T_{D}=\xi_{D}T_{\mathrm{CMB}}$ today. The dark photon does not have a kinetic mixing term with the SM photon, and thus the atomic dark sector only interacts with the SM gravitationally. The bound state of the dark electron and dark proton is generically called dark hydrogen, but does not necessarily imply the SM hydrogen-like limit in which $\mpd \gg \med$. In this work we are concerned with various processes involving this system, particularly those that most strongly impact the cosmological evolution of the atomic dark sector. We now briefly discuss several of these, including Thomson scattering, Bremsstrahlung, collisional bound-bound and bound-free transitions, and finally radiative bound-bound and bound-free transitions. The latter we calculate in detail in Section~\ref{sec:extFieldsinQM}. 

\subsection*{Thomson Scattering and Bremsstrahlung}
Thomson scattering of free dark electrons and dark protons is an important process cosmologically, since it maintains thermal equilibrium between the dark photon and atomic dark matter particles at early times. The Thomson scattering rate is exact, and valid as long as the photon energy is much smaller than the dark fermion mass that is involved in the radiative process. Therefore, as long as we allow both dark electrons \textit{and} dark protons to undergo Thomson scattering this rate is valid even outside the SM-like parameter regime at dark sector temperatures below $\med$.  

Dark electrons and dark protons can undergo Bremsstrahlung. For two point-like massive charges, we typically consider the dominant dipole contribution, where the dipole moment goes like $\boldsymbol{d} = \mu (\frac{e_1}{m_1} - \frac{e_2}{m_2}) \boldsymbol{r}$. For two particles with different masses and opposite charges, we see that the dipole simplifies to $\boldsymbol{d} = \mu \frac{e}{\mu} \boldsymbol{r} = e \boldsymbol{r}$. We can compare this to the case of SM electron-proton Bremsstrahlung in which case the proton degree of freedom drops out and $\mu \simeq m_e$, so that $\boldsymbol{d} = \mu \frac{e}{m_e} \boldsymbol{r} \simeq e \boldsymbol{r}$. We note that dark electron-electron or dark proton-proton Bremsstrahlung can also be considered; however, these processes are always subdominant as the dipole radiation term is strictly zero and the quadrupole term is suppressed by $T/\med$ or $T/\mpd$, respectively. 

From the dipole, one can calculate the dark electron-dark proton Bremsstrahlung cross section and thermally averaged energy loss rate. We note the result from~\cite{Cyr-Racine:2012tfp} that the free-free Bremsstrahlung rate of the two-body system (using reduced mass as opposed to electron mass) from the typical classical description\footnote{The classical description is valid under the following set of assumptions: (a) the electrons are not relativistic, $E_{kin} \ll \med c^2$; (b) the energy of the emitted photon is small compared to the total energy of the electron; and (c) the scattering angle is small such that the acceleration perpendicular to the trajectory is much smaller than the velocity. Each of these approximations are valid within the parameter space of interest.} is accurate for the entire coupling and mass range we consider.

\subsection*{Collisional transitions}

Collisional excitation and ionization play an important role in the dissipative behavior of aDM in galaxies~\cite{Roy:2023zar, Roy:2024bcu}. Here, we explain the assumptions used to calculate rates of collisional processes in the SM-like regime, and highlight how these assumptions may be invalidated for the wider range of aDM parameter space we consider in this work.

In the SM collisional ionization process, the bound atom nucleus is ignored and the bound electron is considered free, but at rest ($v \sim \alpha_{SM} \ll 1$, in units of $c$). As was done by Thomson in 1912, in this case the binary encounter approximation is used~\cite{osti_6676289}, and the scattering cross section can be calculated classically~\cite{Fan:2017}. Importantly, the assumption that the bound fermion can be considered at rest may no longer hold if the coupling $\alphad$ is large. 

Additionally, the rate for collisional ionization and collisional excitation is larger when the incident particle is moving quickly. When the free dark electrons and protons are in thermal equilibrium, their respective velocities scale as $v_e \sim \sqrt{3T/\med}$ and $v_p \sim \sqrt{3T/\mpd}$, which is why in the SM scenario heavier protons are so much slower than the lighter electrons and therefore contribute less to collisional processes. However, as we approach the regime $\mpd \simeq \med$ then $v_p \simeq v_e$. In this case, the dark proton can interact, and it is important to track its degree of freedom throughout the process. 

This generalized three-body interaction involving the bound dark proton, bound dark electron, and the incident particle (dark electron or dark proton) is a much more difficult problem than the radiative processes we will soon detail in this work. The dynamics of the bound state atom are more complicated and a more careful study of collisional processes could be performed to validate the SM scalings for the wider range of mass and couplings we will consider here. These collisional processes are important for late-time cosmological simulations; however, in \autoref{sec:cosmology}, we will not need these collisional processes to study the large-scale cosmological constraints on aDM. 

\subsection*{Radiative transitions}
Finally, we consider radiative bound-bound transitions, recombination, and ionization of dark hydrogen. These two-body processes are more tractable to compute than the collisional ones. We study the cosmological recombination of dark hydrogen and investigate the accuracy of previous results for a wider range of aDM parameter space. Previously used recombination rates are dependent on calculated SM rates, which then are scaled by powers of the dark atomic masses and gauge coupling to extrapolate beyond the SM coupling and masses. Here, we describe the general methodology that will be used in this work, and later we calculate the relevant transition cross sections in detail. 

At early times, the dark photons, dark electrons, and dark protons are in thermal equilibrium, coupled by Thomson scattering, at temperatures above the dark hydrogen binding energy $\bd\equiv\alphad^{2}\mud/2$. The atomic dark sector is fully ionized in this epoch. As the universe expands, the dark sector temperature decreases. When the dark sector temperature falls below the binding energy of dark hydrogen, it becomes energetically preferable for dark electrons and dark protons to form bound states through the process known as recombination. 

In both the SM and atomic dark sector, the evolution of the ionization fraction during recombination is governed by a network of transitions between bound states as well as photo-recombination and photo-ionization. In this work, as in SM recombination calculations, we neglect collisional processes and include only radiative bound-bound and bound-free transitions. State-of-the-art SM recombination calculations use an effective multi-level atom (EMLA) method~\cite{Ali-Haimoud:2010,Ali-Haimoud:2010hou,Lee:2020obi}, which augments the three-level atom formalism of Peebles and Zeldovich et al.~\cite{Peebles:1968ja,1969JETP...28..146Z}, by taking into account transitions between excited states with principal quantum number $n\geq 2$.
Similar to the SM, for aDM we consider `case-B' recombination, in which electrons recombine to $n=2$ states, before radiatively decaying to the $1s$ ground state. 

In the EMLA formalism, the evolution of the ionization fraction $x_{e}$ depends on the effective recombination coefficients $\mathcal{A}_{2\ell}$ and photoionization rates $\mathcal{B}_{2\ell}$, $\ell \in\{s,p\}$~\cite{Ali-Haimoud:2010}. These in turn are computed in terms of recombination coefficients $\alpha_{nl}$ and bound-bound transition rates $R_{nl\rightarrow n_f, l_f}$, for arbitrary $n,\ l,\ n_f,\ l_f$, which are then summed over. These are the quantities that we will compute and compare with rescaled SM values. 
 The recombination coefficient $\alpha_{nl}$ can be calculated from the photo-ionization cross section, $\sigma_{nl \to \text{cont.}}$ (see~\cite{Seager_2000} for details), as follows: 
\begin{align} \label{eqn:recombinationCoef}
    \alpha_{nl}(T_{m},T_{\gamma}) &= 2\left(\frac{n_{nl}}{n_{e} n_{p}}\right)^{LTE} \int_{0}^{\infty} \frac{\sigma_{nl \to \text{cont.}}}{\hbar \omega} \left(\frac{\hbar \omega^3}{2 \pi^2 c^{2}} + B(\omega, T_{\gamma})\right) e^{-\frac{\hbar \omega}{k_{b} T_{m}}} d \omega \,.
\end{align}
Here, $n_e$ ($n_p$) is the number density of dark electrons (dark protons), and $n_{n,l}$ is the number density of the dark bound state, with quantum numbers $n,l$, in local thermal equilibrium. The spectral radiance of photons with energy $\omega$ is $B(\omega,T_{\gamma})~=~\frac{\hbar \omega^3}{4 \pi^3 c^2}\frac{1}{e^{\frac{\hbar \omega}{k_b T_\gamma}}-1}$. The bound-bound transition rate for atoms in a thermal bath of photons is~\cite{Seager_2000}
\begin{align} \label{eqn:bbRate}
    R_{n l \to n_f l_f} &=  n_{nl} \int^{\infty}_0 d\omega \frac{4\pi}{\hbar \omega} B(\omega, T_{\gamma})\sigma_{nl \rightarrow n_f l_f} \,,
\end{align}
where $\sigma_{nl \rightarrow n_f l_f}$ is the cross section for transitions from state $(n,l)$ to $(n_{f}, l_{f})$. 
Next, we calculate the photo-ionization cross section $\sigma_{nl \rightarrow \mathrm{cont.}}$ and bound-bound cross section $\sigma_{nl \rightarrow n_f l_f}$ without relying on the approximations $\med\ll \mpd$ and $\alphad\ll 1$. 
\section{Solutions of the dark hydrogen Schr\"odinger Equation} \label{sec:QM}
\subsection*{Dark hydrogen Hamiltonian}
In order to compute the bound-bound and bound-free cross sections for dark hydrogen, we first calculate the continuum and bound states that solve the two-body Schr\"odinger equation. We later present the photon interaction Hamiltonian and compute the cross sections in terms of transition amplitudes between these states. Although the general two-body states have been found previously~\cite{Burgess:1960, Burgess1965}, we set up the Hamiltonian and quote the two-body bound and continuum states for completeness.

The general two-body Hamiltonian for a dark electron and dark proton with a Coulomb interaction described by the non-relativistic potential $V(r) = - \frac{\hbar c \alphad}{r}$ is given by 
\begin{align} \label{eqn:FreeHamiltonian_rel}
    H_{0} = \frac{\mathbf{p}_{p}^{2}}{2 \mpd} + \frac{\mathbf{p}_{e}^{2}}{2 \med} - \frac{\hbar c\alphad}{|\mathbf{r}_{e}-\mathbf{r}_{p}|}\ ,
\end{align}
where $\mathbf{p}_{e}$ ($\mathbf{p}_{p}$) is the three-momentum of the dark electron (proton) and $\mathbf{r}_{e}$ ($\mathbf{r}_{p}$) is the position of the dark electron (proton); note that we drop the $D$ subscript for these coordinates. In general, there are atomic corrections, which are small in the SM due to $\alpha_\SM \ll 1$. We determine the parameter ranges in which these can continue to be neglected for the purpose of computing the cross sections relevant for recombination. We note that the binding energy of the dark hydrogen atom scales like the Rydberg energy, or $E \sim \alphad^2 \mud$. Spin-orbit coupling induces corrections to the kinetic energy that scale as $\Delta E_{\mathrm{fine}} \sim \alphad^4 \mud$. Hyperfine magnetic dipole interactions between the dark electron and dark proton spin can induce further corrections to this energy $\Delta E_{\mathrm{hyperfine}} \sim \alphad^4 \mud (\mud / \mpd)$, for $\med \leq \mpd$ (note that for positronium, fine structure and hyperfine structure energy corrections are of the same magnitude). Further corrections, such as the Lamb shift from QED dark electron self-energy and vacuum polarization ($\Delta E_{\mathrm{Lamb}} \sim \alphad^5 \mud $), scale with higher powers of $\alphad$. 
For large enough $\alphad$, these terms, which are usually sub-dominant in the SM, may become comparable with the binding energy. Therefore, we limit our analysis to $\alphad \lesssim 0.3$ where these effects remain sub-dominant. Further studies of $0.3 < \alphad < 1$ would require a careful analysis of relativistic corrections as well as the fine structure of hydrogen and other higher-order interactions. We note that varying the dark fermion masses from the hydrogen-like to positronium-like limit does not strongly enhance these sub-dominant effects, allowing us to consider arbitrary $\med/\mpd$ ratios. 

The eigenvalues and eigenfunctions of this Hamiltonian can be solved exactly by transforming to the center-of-mass (COM) coordinates. We define the COM (upper case letters) and the relative (lower case letters) position and momentum coordinates by 
\begin{align} \label{eqn:COMandRELCoord}
\begin{aligned}
    \mathbf{R} &= \frac{\mpd \mathbf{r}_{p} + \med \mathbf{r}_{e}}{\mpd+\med} \ , \\
    \mathbf{P} &= (\mpd+\med) \frac{d\mathbf{R}}{dt} \ ,
\end{aligned}
\hspace{2cm}
\begin{aligned} 
    \mathbf{r} &= \mathbf{r}_{e} - \mathbf{r}_{p} \ , \\
    \mathbf{p} &= \frac{\mpd\med}{\mpd+\med} \frac{d\mathbf{r}}{dt}\ .
\end{aligned}
\end{align}
It is also convenient to define the reduced mass ($\mud$) and total mass ($\mHd$) of the dark hydrogen bound state as
\begin{align}
\begin{aligned}
    \mud &= \frac{\mpd \med}{\mpd+\med}\ ,
\end{aligned}
\hspace{2.5cm}
\begin{aligned}
    \mHd &= \mpd+\med \ .
\end{aligned}
\end{align}
We can then write the Hamiltonian from~\cref{eqn:FreeHamiltonian_rel} in terms of the COM and relative coordinates
\begin{align} \label{eqn:FreeHamiltonian_COM}
    H_{0} = \frac{\mathbf{P}^{2}}{2 \mHd} + \frac{\mathbf{p}^{2}}{2 \mud} - \frac{\hbar c \alphad}{r} \ .
\end{align}

\subsection*{Dark hydrogen wave functions} 
Since the Hamiltonian in~\cref{eqn:FreeHamiltonian_COM} is separable, the solution can be decomposed into two eigenstates, $ \ket{\Psi} = \ket{\Psi_{\text{COM}}} \otimes \ket{\Psi_\text{rel}}$ and the COM eigenstates $\ket{\Psi_{\text{COM}}}$ can be calculated independently from the relative eigenstates $\ket{\Psi_\text{rel}}$, 
\begin{align}\label{eqn:com_hamiltonian}
    \frac{\mathbf{P}^{2}}{2 \mHd} \ket{\Psi_{\text{COM}}} = E_{\text{COM}} \ket{\Psi_{\text{COM}}} \ ,
\end{align}
\begin{align} \label{eqn:rel_hamiltonian}
    \left(\frac{\mathbf{p}^{2}}{2 \mud} - \frac{\hbar c \alphad}{r}\right)\ket{\Psi_{\text{rel}}} = E_{\text{rel}} \ket{\Psi_{\text{rel}}} \ .
\end{align}
The solutions to~\cref{eqn:com_hamiltonian} are momentum eigenstates ($\ket{\Psi_{\text{COM}}} = \ket{\textbf{P}}$) with positive energy $E_{\text{COM}} = \frac{\textbf{P}^2}{2 \mHd}$, where $\textbf{P} \in \mathbb{R}^{3}$.
\cref{eqn:rel_hamiltonian} admits two classes of solutions, bound states that correspond to negative energies ($E_{\text{rel}} < 0$) and continuum states that correspond to positive energies ($E_{\text{rel}} \geq 0$). We denote the bound wavefunctions by $\psi_{nlm}(r, \theta, \phi) \equiv  \braketm{r, \theta, \phi}{nlm}$, where $n, l, m \in \mathbb{Z}$, $n>0$, $0\leq l < n$, and $-l\leq m \leq l$, and we denote the continuum wavefunctions by $\psi_{klm}(r, \theta, \phi) \equiv \braketm{r, \theta, \phi}{klm}$, where $k \in \mathbb{R}$, $l, m \in \mathbb{Z}$, $k,l \geq 0$, and  $-l\leq m \leq l$.  Therefore, the total wavefunction is given by 
\begin{align} \label{eqn:totalWaveFunction}
    & \ket{\mathbf{R},\mathbf{r}}{\Psi} = \ket{\mathbf{P}} \otimes \begin{cases} 
        \psi_{nlm}(r,\theta,\phi) & E_{\text{rel}} < 0 \\
        \psi_{klm}(r,\theta,\phi)  & E_{\text{rel}} \geq 0 \\
        \end{cases}  \ .
\end{align}
The bound and continuum wavefunctions and their corresponding energies can be calculated exactly. Written in terms of the confluent hypergeometric function $_{1} F _{1}(a,b,x)$ (see derivations in~\cite{landau1981quantum, sakurai2017modern, Burgess:1960, Burgess1965}), the bound wavefunction is given by 
\begin{align}
\label{eqn:psinlm}
    \psi_{nlm}(r,\theta,\phi)
    &= N_{b} P_{l}^{m}(\cos \theta) e^{i m \phi} r^{l}  e^{-\frac{r}{a_{o} n}} \left._{1}F_{1}\right.\left(l+1-n; 2l+2; \frac{2 r}{a_{o} n}\right) \ ,
\end{align}
and the continuum wavefunction is given by 
\begin{align}
    \psi_{klm}(r,\theta,\phi) 
     &= N_{c}(a_0 n k) P_{l}^{m}(\cos \theta) e^{i m \phi} r^{l}e^{ikr} \left._{1} F _{1}\right.(l+1 - \frac{i}{a_{o} k} ; 2l + 2; -2ikr) \ ,
\end{align}
where $a_0 = \frac{\hbar}{\mud \alphad c}$ is the Bohr radius. $N_{b}$ and $N_{c}(a_0 n k)$ are the normalizations for bound and continuum states, respectively, 
\begin{align} \label{eqn:N_b}
    N_{b} &= \frac{2^{l}}{a_{o}^{l+\frac{3}{2}} n^{l+2}\sqrt{\pi}}\sqrt{\frac{(l-m)!(n+l)!}{(l+m)! (2l+1)! (2l)! (n-l-1)!}} \ ,
\end{align}
and 
\begin{align} \label{eqn:N_c}
    N_{c}(k') &= \frac{2^{l-\frac{1}{2}}}{a_{o}^{l+1} n^{l+\frac{1}{2}}\sqrt{\pi}}\sqrt{\frac{(l-m)! k' \left(1+\coth\left(\frac{n\pi}{ k' }\right)\right)\prod_{s=0}^{l} \left(n^2+\left(k' s\right)^{2}\right)}{(l+m)! (2l+1)! (2l)!}} \ .
\end{align}
The energies for the bound and continuum wavefunctions are
\begin{align}\label{eqn:relenergies}
    E_{\text{rel}} = \begin{cases} 
        - \frac{\mud c^2 \alphad^{2}}{2 n^2} & \text{  bound}\\
        \frac{\hbar^{2} k^{2}}{2 \mud}  &\text{  continuum}\\
     \end{cases} \ ,
\end{align}
so the total energy of the aDM system is $E = E_{\text{COM}} + E_{\text{rel}}$.

\section{Computing radiative transition cross sections} \label{sec:extFieldsinQM}
We are interested in calculating transition rates between the bound to bound and bound to continuum states calculated in \autoref{sec:QM}. In the early universe, dark atoms exist within a thermal background of dark photons that can interact and mediate atomic transitions. We now treat these interactions in perturbation theory. We assume that the matter particles have a negligible effect on the dark photon field and that the density of dark photons is large enough such that it can be approximated by a classical field, described by the electromagnetic four-potential $A^{\nu} = (A^0, \mathbf{A})$. Therefore, the full non-relativistic Hamiltonian is given by
\begin{align}\label{eqn:starting_interaction_hamiltonian}
    H = \frac{\left(\mathbf{p}_{p} - e_D \mathbf{A}(\mathbf{r}_{p})\right)^{2}}{2 \mpd} + \frac{\left(\mathbf{p}_{e} + e_D \mathbf{A}(\mathbf{r}_{e})\right)^{2}}{2 \med} + e_D \left(A^{0}(\mathbf{r}_{p}) - A^{0}(\mathbf{r}_{e})\right) - \frac{\hbar c \alphad}{|\mathbf{r}_{e}-\mathbf{r}_{p}|} \ .
\end{align}
We impose the Coulomb gauge ($\nabla \cdot \mathbf{A} = 0$) so that the dark matter momentum and the dark photon field commute.
We can rewrite~\cref{eqn:starting_interaction_hamiltonian} in terms of the exactly solvable Hamiltonian in~\cref{eqn:FreeHamiltonian_COM} and a small time-dependent perturbation $V(t)$,
\begin{align}
    H &= H_{0} + V(t) \ , 
\end{align}     
where 
\begin{align} \label{eqn:potential}
    V(t)= e_D \left(\frac{\mathbf{A}(\mathbf{r}_{e})\cdot \mathbf{p}_{e}}{\med}-\frac{\mathbf{A}(\mathbf{r}_{p})\cdot \mathbf{p}_{p}}{\mpd} + A^{0}(\mathbf{r}_{p}) - A^{0}(\mathbf{r}_{e})\right)+ \frac{e_D^2}{2}\left(\frac{\mathbf{A}^{2}(\mathbf{r}_{e})}{\med}+\frac{\mathbf{A}^{2}(\mathbf{r}_{p})}{\mpd}\right) \ .
\end{align} 
The four-vector potential describing a monochromatic wave in Coulomb gauge is
\begin{align}
    A^{\mu} (\boldsymbol{r}) &= \left(0, A_{o} \hat{\epsilon} \left[ e^{i \frac{\omega}{c} \hat{n}\cdot \boldsymbol{r} - i \omega t} + e^{-i \frac{\omega}{c} \hat{n}\cdot \boldsymbol{r} + i \omega t }\right]\right),
\end{align} 
where $A_0$ is the amplitude, $\hat{\epsilon}$ is the polarization direction, $\omega$ is the frequency, $c$ is the speed of light, $\hat{n}$ is the propagation direction, and $\boldsymbol{r}$ is the position. Since the field strength and $e^2_D (\equiv 4 \pi \alpha_D)$ are small, we can neglect the terms quadratic in $\mathbf{A}$. We change variables to use the COM and relative coordinates, and choose a coordinate system such that $\hat{\epsilon} \parallel \hat{x}$ and $\hat{n} \parallel \hat{z}$. We can then write the potential as $V(t) = \tilde{V}(\omega) e^{i\omega t} + \tilde{V}^{\dagger}(\omega) e^{-i \omega t} = \tilde{V}(\omega) e^{i\omega t} + \tilde{V}(-\omega) e^{-i \omega t}$, where we have defined   
\begin{align}
    \tilde{V}(\omega) &= A_0 e_D e^{-i \frac{\omega Z}{c}} \left[\left(e^{-i\frac{\omega \mud z}{\med c}}-e^{i\frac{\omega \mud z}{\mpd c}}\right)\frac{P_{x}}{\mHd}   +  \left(\frac{e^{-i\frac{\omega \mud z}{\med c}}}{\med}+\frac{e^{i\frac{\omega z}{\mpd c}}}{\mpd}\right)p_{x}\right] \: .
\end{align}
Clearly, the dark proton and dark electron experience a time-dependent harmonic potential due to the dark photon field. 

Using Fermi's Golden Rule for harmonic potentials, we calculate the transition rate between an arbitrary initial ($\{\mathbf{P}^{i},i\}$) and final ($\{\mathbf{P}^{f},f\}$) state, where $i$ and $f$ label either bound or continuum states of the relative degree of freedom, 
\begin{align} \label{eqn:FermiGoldenRule}
    \Gamma_{\{\mathbf{P}^{i},i\} \to \{\mathbf{P}^{f},f\}} &=\frac{2 \pi}{\hbar}\delta\left(E^{f}_{\text{rel}}-E^{i}_{\text{rel}}+\frac{(\textbf{P}^f)^2 - (\textbf{P}^i)^2}{2 \mHd}+\hbar \omega\right) \left|\bra{\mathbf{P}^{f},f} \tilde{V} \ket{\mathbf{P}^{i},i}\right|^2  + \left(\omega\to-\omega\right) \ .
\end{align} 
The relative energies are defined in~\cref{eqn:relenergies}. We note that both $\tilde{V}$ and the wavefunctions are separable with respect to COM and relative coordinates. Therefore, the transition amplitude in~\cref{eqn:FermiGoldenRule} can be split into two parts, one which depends only on the COM degree of freedom, and another that only depends on the relative degree of freedom. To do so, we have trivially acted on the COM initial momentum state to get its eigenvalue ($P_x\ket{\mathbf{P}^{i}} = P^i_x \ket{\mathbf{P}^{i}}$) and separate terms,
\begin{align} \label{eqn:generalOverlap}
\bra{\mathbf{P}^{f},f} \tilde{V} (\omega) \ket{\mathbf{P}^{i},i} & = A_0 e_D  \bra{\mathbf{P}^{f}} e^{-i \frac{\omega Z}{c}} \ket{\mathbf{P}^{i}} \nonumber \\ 
& \times \bra{f}\left[\left[e^{-i\frac{\omega \mud z}{\med c}}-e^{i\frac{\omega \mud z}{\mpd c}}\right]\frac{P_{x}^{i}}{\mHd} +  \left[\frac{e^{-i\frac{\omega \mud z}{\med c}}}{\med}+\frac{e^{i\frac{\omega \mud z}{\mpd c}}}{\mpd}\right]p_{x}\right]\ket{i} \ .
\end{align}
We normalize the COM coordinates such that $\braket{\mathbf{R} | \mathbf{P}} = V^{-1/2}e^{i\frac{\mathbf{P}\cdot \mathbf{R}}{\hbar}}$ and solve for the COM momentum overlap,
\begin{align}
    \bra{\mathbf{P}^{f}} e^{-i \frac{\omega Z}{c}} \ket{\mathbf{P}^{i}} = \frac{(2 \pi \hbar)^{3}}{V}\delta \left(P^{f}_{x} - P^{i}_{x}\right)\delta \left(P^{f}_{y} - P^{i}_{y}\right) \delta \left(P^{f}_{z} - P^{i}_{z} + \frac{\hbar \omega}{c}\right). 
\end{align}
At this point, it is helpful to identify that we only need to calculate bound-bound and bound-continuum cross sections. Therefore, we can set the initial state to $\ket{i} = \ket{nlm}$ and allow for general final $\bra{f}$ states. The transition amplitudes that must be evaluated are: 
\begin{align} \label{eqn:overlapS1}
    S_{1f} &= \bra{f}\left(e^{-i\frac{\omega \mud z}{\med c}}-e^{i\frac{\omega \mud z}{\mpd c}}\right)\ket{nlm} \ , \\
    \label{eqn:overlapS2}
    S_{2f} &= \bra{f}\left(\frac{1}{\med}e^{-i\frac{\omega \mud z}{\med c}}+\frac{1}{\mpd}e^{i\frac{\omega \mud z}{\mpd c}}\right) p_x\ket{nlm} \ ,
\end{align}
where we have trivially factored out the COM momentum $P^i_x/\mHd$ to define $S_{1f}$ with only relative degrees of freedom. Detailed calculations of~\cref{eqn:overlapS1} and~\cref{eqn:overlapS2} are shown in~\autoref{sec:overlaps}. We can re-write~\cref{eqn:generalOverlap} in terms of these overlaps as, 
\begin{align} \label{eqn:TotalOverlap}
\bra{\mathbf{P}^{f},f} \tilde{V}(\omega) \ket{\mathbf{P}^{i},nlm} &= A_0 e_D \frac{(2 \pi \hbar)^{3}}{V} \delta \left(P^{f}_{\perp} - P^{i}_{\perp}\right) \delta \left(P^{f}_{z} - P^{i}_{z} + \frac{\hbar \omega}{c}\right) \left(S_{1f}\frac{P_{x}^{i}}{\mHd} + S_{2f}\right) \ ,
\end{align}
where $\delta(P^{f}_{\perp} - P^{i}_{\perp}) \equiv \delta (P^{f}_{x} - P^{i}_{x})\delta (P^{f}_{y} - P^{i}_{y})$. We substitute the overlap into the transition rate in~\cref{eqn:FermiGoldenRule} and divide by the photon flux to obtain a cross section
\begin{align}
\label{eqn:rate_to_xs}
    \sigma_{\{\mathbf{P}^{i},nlm\} \to \{\mathbf{P}^f,f\}} = \frac{\hbar}{2c \epsilon_{0} \omega A_{0}^{2}}\times\Gamma_{\{\mathbf{P}^{i},nlm\} \to \{\mathbf{P}^f,f\}} \ .
\end{align}
 We now integrate over the total final-state COM phase space ($d^3R^f d^3P^f$), because we are interested in the total cross section for transitions between different $\ket{\Psi_\text{rel}}$. The integral over COM position space is trivial because the cross section is independent of $\textbf{R}^f$. The remaining integrals over COM momentum can be completed using three of the six COM Dirac delta functions that come from squaring~\cref{eqn:TotalOverlap}. The remaining three Dirac delta functions, which are evaluated at zero, can be identified with $\frac{V}{(2 \pi \hbar)^{3}}$. All factors of volume cancel, and we are left with 
\begin{align}
\sigma_{\{\mathbf{P}^{i},nlm\} \to \{f\}} =& \frac{4 \pi^2 \hbar\alphad}{\omega}\left|S_{1f}\frac{P_{x}^{i}}{\mHd}   + S_{2f}\right|^{2} \delta(E^{f}_{rel}-E^{i}_{rel}+\frac{\hbar^{2} \omega^{2}}{2c^{2}\mHd} - \frac{P^{i}_{z} \hbar \omega}{c \mHd}+\hbar \omega) \nonumber \\
& +\left(\omega \to -\omega\right) \ . 
\end{align} 
We assume that the dark fermions in the early universe have a Maxwell-Boltzmann momentum distributions at a common matter temperature, and therefore we average over the initial COM momentum,
\begin{align}
    \sigma_{nlm \to \{f\}} = \int_{-\infty}^{\infty} d^{3}P^{i}\frac{e^{- \frac{(\mathbf{P}^{i})^{2}}{2 \mHd k_{B} T_m}}}{(2 \pi \mHd k_{B} T_{m})^{3/2}} \sigma_{\{\mathbf{P}^{i},nlm\} \to \{f\}} \ .
\end{align} 
Performing the three-dimensional integral, we find
\begin{align}
\label{eq:sigma_rel}
\sigma_{nlm \to \{f\}} =& \frac{4 \pi^2 c \alphad \mHd^{1/2} \left(\frac{k_{B} T_m}{\mHd}|S_{1f}|^{2} + |S_{2f}|^{2}\right)}{ \omega^2 (2 \pi k_{B} T_m)^{1/2}}e^{-\frac{\mHd c^2}{2 k_B T_m \hbar^2 \omega^2} \left(E_{\text{rel}}^f - E_{\text{rel}}^i + \frac{\hbar^2 \omega^2}{2 \mHd c^2} + \hbar \omega \right)^2} \nonumber \\
& + \left(\omega \to -\omega\right)\ . 
\end{align} 
The final state in this cross section is general, allowing for $f$ to be a bound state $\ket{n_f l_f m_f}$ or continuum state $\ket{k_f l_f m_f}$. To compute the cross section to transition to a bound state defined by $\ket{n_{f} l_{f}}$, we now average over $m\in \{-l,\dots,l\}$, sum over $m_{f}\in\{-l_{f},\dots,l_{f}\}$, and use $E_{\text{rel}}^{f}-E_{\text{rel}}^{i}=\frac{\alphad^{2}\mud}{2}\left(\frac{1}{n_{i}^{2}} - \frac{1}{n_{f}^{2}}\right)$ to obtain a final cross-section ${\sigma_{nl\rightarrow n_{f} l_{f}}\equiv\frac{1}{2l+1}\sum_{m,m_{f}}\sigma_{nlm\rightarrow n_{f}l_{f}m_{f}}}$. In general, the bound-bound cross sections for the transitions $(n l) \rightarrow (n_f l_f)$ and $(n_f l_f) \rightarrow (n l)$ are not equal because of the COM degree of freedom that is integrated out. This is different from the SM description, where the COM degree of freedom is neglected. See~\autoref{sec:DeviationsFromSM} for a more detailed discussion. 
For transitions to the continuum, we average over $m$ and sum over $m_{f}$ and $l_{f} \in \{0, \dots, \infty \}$ as well as integrate over $k$. The difference in energy levels is $E_{\text{rel}}^{f}-E_{\text{rel}}^{i}=\frac{k^{2}}{2\mud}+\frac{\alphad^{2}\mud}{2n_{i}^{2}}$. Therefore, the photo-ionization cross section is  ${\sigma_{nl\rightarrow\textrm{cont.}}\equiv\frac{1}{2l+1}\sum_{m,m_{f},l_{f}}\int dk\ \sigma_{nlm\rightarrow kl_{f}m_{f}}}$. 
Although the sum over $l_f$ is over all values, we note that terms with $|l-l_f| > 1$ will be suppressed compared to the dominant $l-l_f = \pm 1$ terms. We find that summing over the dominant $\Delta l = \pm 1$ terms is sufficient for convergence, as is also true in the SM. The recombination coefficient $\alpha_{nl}$ is 
\begin{equation}
    \label{eqn:recombinationCoef2}
    \alpha_{nl} = \frac{2 \hbar^3}{c^2} \sqrt{\frac{2}{\pi}}  (2l+1)  \left(\frac{1}{k_b  T_{m}\mud}\right)^{3/2} e^{\frac{E_{n}}{k_b T_{m}}}\int_{0}^{\infty}d\omega \frac{e^{(\frac{1}{k_b  T_{\gamma}}-\frac{1}{k_b T_{m}})\hbar \omega}}{e^{\frac{\hbar \omega}{k_b T_{\gamma}}}-1}  \omega^{2}\sigma_{nl \to \text{cont.}} \ ,
\end{equation}
where we have used the Saha equation to re-write the number densities in~\cref{eqn:recombinationCoef}, and the bound-bound transition rate $R_{n l \to n_f l_f}$ is
\begin{equation}
    \label{eqn:bbRate2}
    R_{n l \to n_f l_f} = \frac{n_{nl}}{ c^2 \pi^2}\int_{0}^{\infty} d\omega \omega^2\frac{\sigma_{nl\rightarrow n_{f}l_{f}}}{e^{\frac{\hbar \omega}{k_b T_{\gamma}}}-1} \ .
\end{equation}

\section{Comparison of rates calculated from first-principles with Standard Model scaling} \label{sec:CompareSMScalings}
In this section, we show that our first-principles calculations reproduce the anticipated SM recombination and bound-bound transition rates to within sub-percent accuracy at $\alpha_D = \alpha_\SM$ and $\mpd \gg \med$. We note that beyond the dipole approximation (which has selection rules $\Delta l = \pm 1$), different selection rules can apply and these rates can also be calculated using our first-principles approach. We confirm that the $\Delta l \ne \pm 1$ transition rates are negligibly small compared to the dominant $\Delta l = \pm 1$ transitions. For the remainder of this section, we compare our generalized results (namely,~\cref{eqn:recombinationCoef2} and~\cref{eqn:bbRate2}) to the SM scalings in a variety of scenarios. It is useful to parameterize the dark photon temperature in terms of its ratio to the energy of the initial bound state with principal quantum number $n$, $\mathcal{R}_{\gamma}\equiv T_{\gamma}/E_{n}$, and the atomic dark matter temperature in terms of its ratio to the dark photon temperature $\mathcal{R}_{m}\equiv T_{m}/T_{\gamma}$.

\subsection*{Continuum-Bound Transitions}
\begin{figure} [t!]
    \centering
    \includegraphics[width=0.50 \linewidth]{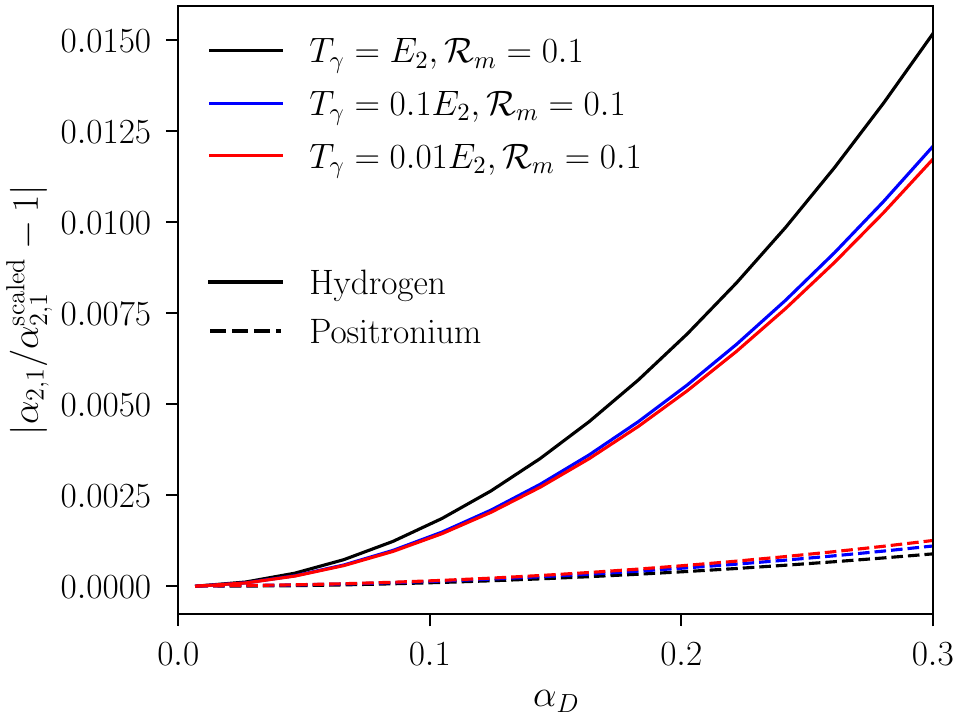}
    \caption{Fractional difference between the first-principles calculation of the recombination coefficient $\alpha_{21}$ and the value obtained from scaling the SM result in \cref{eqn:alpha-scaled} when setting the dark electron and dark proton masses to their SM values (\textbf{solid lines}) and when setting them both equal to SM electron mass to obtain dark positronium (\textbf{dashed lines}).  We show three different temperatures: $T_{\gamma} = E_n$ (\textbf{black}), $T_{\gamma} = 0.1 E_n$ (\textbf{blue}), and $T_{\gamma} = 0.01 E_n$ (\textbf{red}). In all cases, $E_n = \bd / n^2$ with $n=2$ and the matter-radiation ratio is $\mathcal{R}_{m} = 0.1$. We note that for $\mathcal{R}_{m} = 1.0$ these deviations can be as high as 3\% for $\alphad\simeq 0.3$. \label{fig:recombination_difference} } 
\end{figure}

The usual SM calculation of the recombination coefficient $\alpha_{nl}$ yields the following scaling with couplings and masses~\cite{Ali-Haimoud:2010, Cyr-Racine:2012tfp, Fan:2017}:
\begin{equation}\label{eqn:alpha-scaled}
\alpha^{\rm{scaled}}_{nl} (\alphad, \mud) = \alpha_{nl} (\alpha_\SM,\mu_{\SM}) \left( \frac{\alphad}{\alpha_{\SM}} \right)^3\left( \frac{\mu_{\SM}}{\mud} \right)^{3/2} \left( \frac{T^{\SM}_m}{T_m} \right)^{1/2}. 
\end{equation}
Since $T_m \equiv \mathcal{R}_m \mathcal{R}_\gamma E_n \propto \alphad^2 \mud/n^2$, for fixed $\mathcal{R}_m$ and $\mathcal{R}_\gamma$ the recombination coefficient scales as $\alphad^2/ \mud^2$.  
We consider the most important recombination transition in the cosmological context, $\alpha_{21}$. In~\autoref{fig:recombination_difference}, we show the fractional difference between the SM scaling and our general result as a function of the coupling at three different choices of dark photon and dark matter temperatures (for each curve, $\mathcal{R}_m$ and $\mathcal{R}_\gamma$ are held fixed), which explore the relevant temperature range for recombination. We examine in detail two mass scenarios, hydrogen-like with $\med/\mpd \sim 10^{-3}$ and positronium-like with $\med=\mpd$. As the coupling increases to 0.3, the largest deviations from SM scalings are for hydrogen-like masses at higher radiation temperatures. Even in the most extreme case, the SM-scalings are accurate to within 2 \% for the temperatures we have shown.\footnote{We note, that slightly larger deviations, $\mathcal{O}(3\%)$, occur for $\mathcal{R}_m = 1.0$. However, in the case of $T_m = T_\gamma$ there is a cancellation in the exponential term that makes this particular temperature a poor general example. }

\subsection*{Bound-Bound Transitions}
The SM scaling relation for bound-bound transition rates is~\cite{Ali-Haimoud:2010}
\begin{equation} \label{eqn:SM_scaling}
R^{\rm{scaled}}_{n l \to n_f l_f} (\alphad, \mud) = R_{n l \to n_f l_f} (\alpha_\SM, \mu_{\SM})\left( \frac{\alphad}{\alpha_\SM} \right)^5 \left( \frac{\mud}{\mu_{\SM}} \right) \ .
\end{equation}
The SM bound-bound rates, $R_{n l \to n_f l_f} (\alpha_{\SM}, \mu_{\SM})$, can be calculated accurately using the dipole approximation. 

\begin{figure}[t!]
    \centering
    \begin{minipage}[t]{0.49\linewidth}
        \centering
        \includegraphics[width=\linewidth]{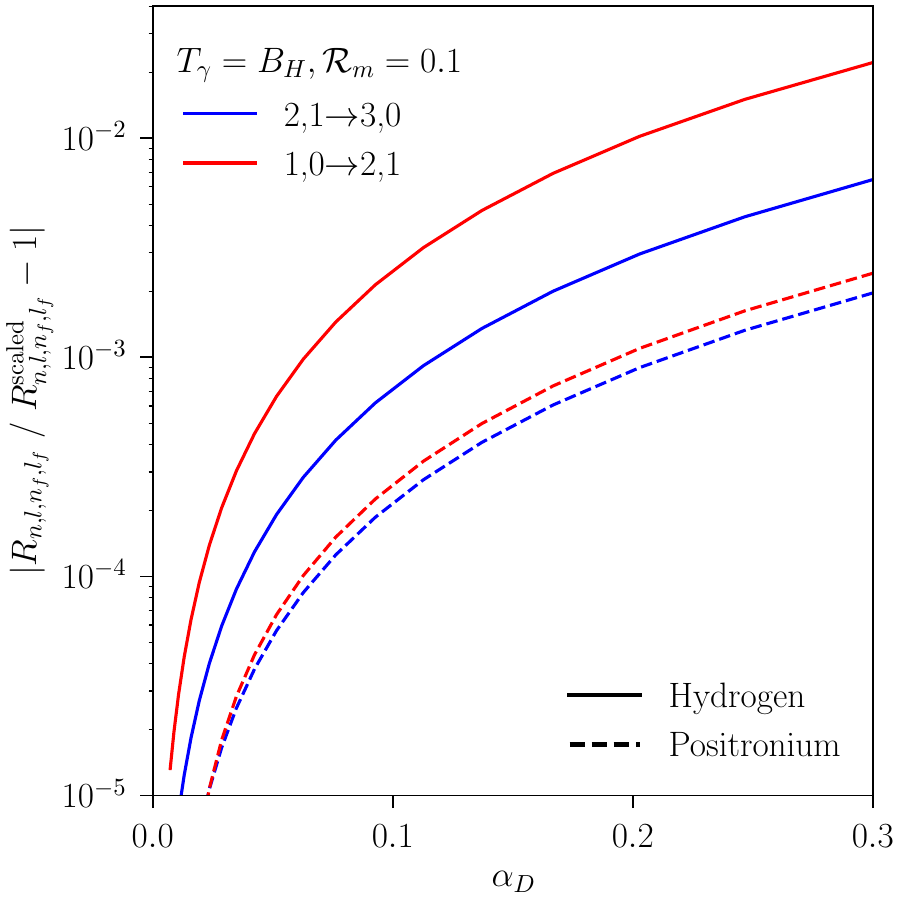}
    \end{minipage}
    \hfill
    \begin{minipage}[t]{0.49\linewidth}
        \includegraphics[width=\linewidth]{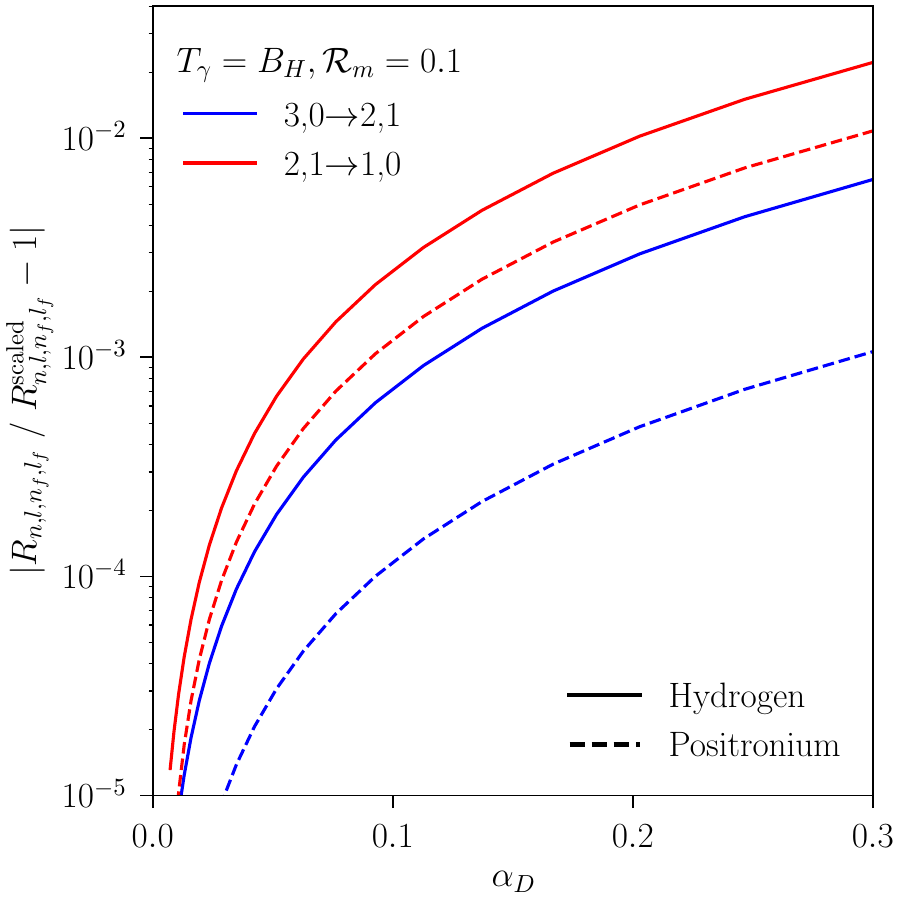}
    \end{minipage}
    \caption[]{Fractional difference between the first-principles calculation of the bound-bound transition rate to the value obtained from scaling the SM result in \cref{eqn:SM_scaling} versus $\alphad$ at radiation temperature $T_\gamma = \bd = \alphad^2 \mud/2$ and a matter-radiation temperature ratio of $\mathcal{R}_{m} = 0.1$, when setting the dark electron and dark proton masses to their SM values (\textbf{solid lines}), and when setting them both equal to SM electron mass to obtain positronium (\textbf{dashed lines}). We consider two examples of photo-absorption processes, specifically $2,1 \rightarrow 3,0$ and $1,0 \rightarrow 2,1$ (\textbf{left}) as well as the stimulated emission processes, specifically $ 3,0 \rightarrow 2,1$ and $2,1 \rightarrow 1,0$ (\textbf{right}). }
    \label{fig:R_ratio}
\end{figure}

We begin by discussing transition rates as a function of the coupling $\alphad$, shown in~\autoref{fig:R_ratio} at radiation temperature $T_\gamma = \bd$ and matter-radiation temperature ratio $\mathcal{R}_{m} = 0.1$. We show four illustrative processes, two photo-absorption transitions 
($2,1 \rightarrow 3,0$ and $1,0 \rightarrow 2,1$) and the symmetric photo-emission transitions ($3,0 \rightarrow 2,1$ and $2,1 \rightarrow 1,0 $). As the coupling $\alphad$ grows, deviations from SM scalings become larger. Although the examples shown in~\autoref{fig:R_ratio} have larger deviations in hydrogen compared to the related transition for positronium, this pattern need not hold for all transitions or at all temperatures.  

\begin{figure}[t!]
    \centering
    \includegraphics[width=0.9 \linewidth]{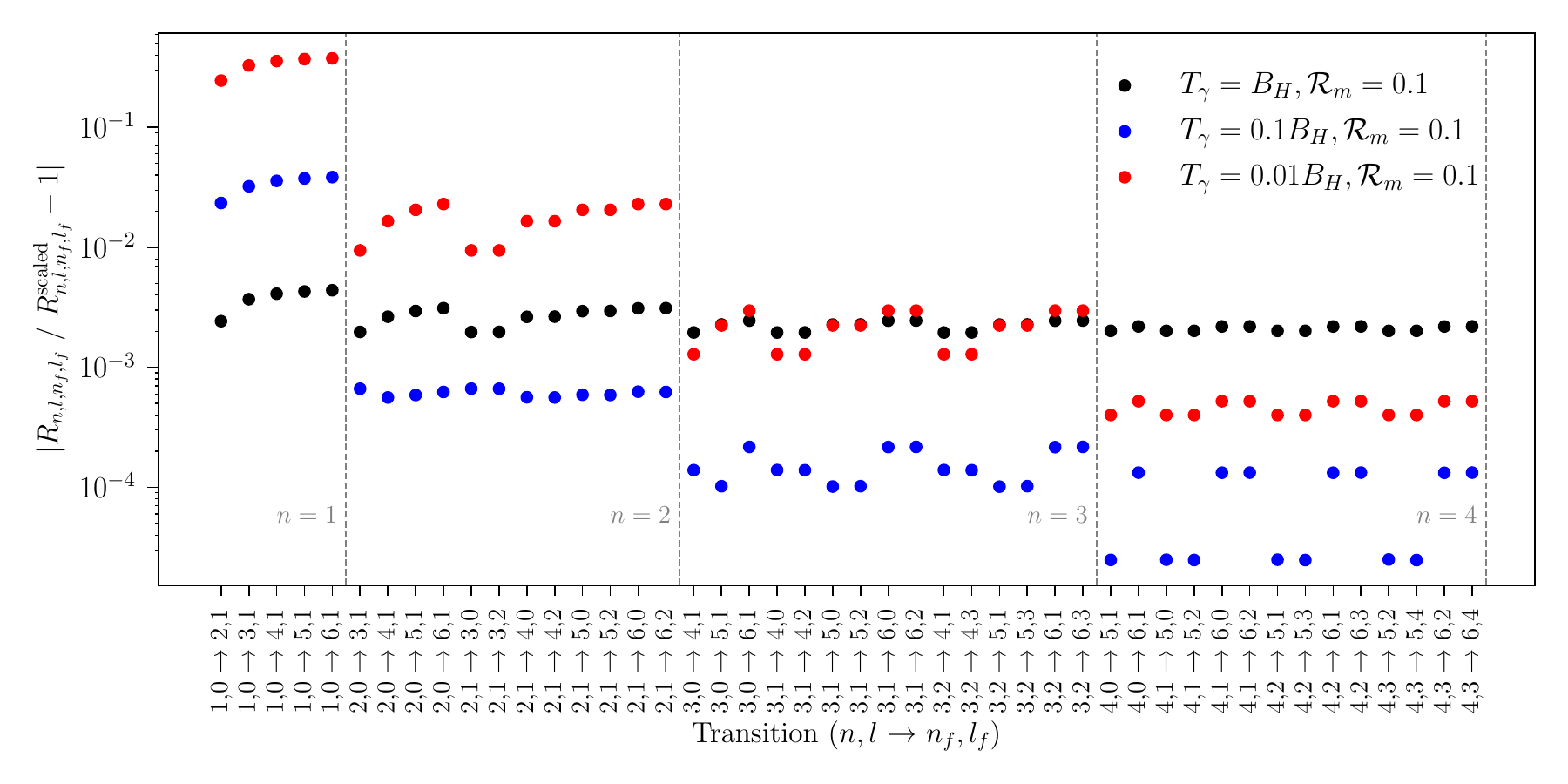}
    \\
    \includegraphics[width=0.9 \linewidth]{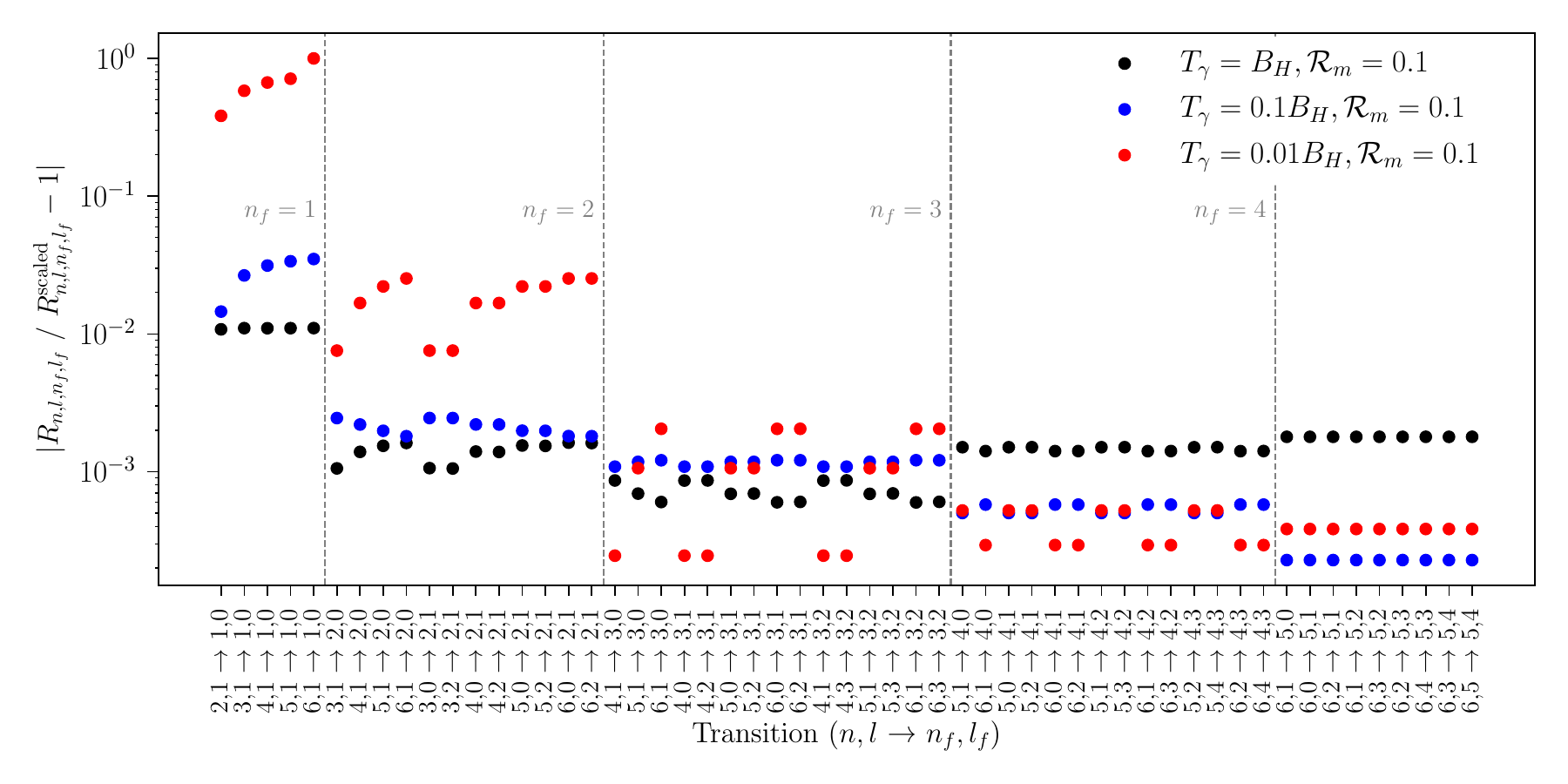}
    \caption[]{Fractional difference between the bound-bound transition rates and the value obtained from scaling the SM result for positronium masses and large coupling ($\alphad=0.3$). We show many transitions at three different physical temperatures, with the radiation temperature always fixed at $T_\gamma \propto \bd = \alphad^2 \mud/2$. We show absorption transitions up to $n_f =6$ (\textbf{top}) as well as emission processes up to $n=6$ (\textbf{bottom}). Transitions to the ground state for emission show the most significant deviations of $\sim \mathcal{O}(1)$.
    }
    \label{fig:converged_percent_difference}
\end{figure}

We study a larger set of transitions at $\alphad=0.3$ (which will highlight the largest deviations) for three different radiation temperatures, shown in~\autoref{fig:converged_percent_difference}. We focus on the SM positronium limit; however, the same analysis can trivially be shown at any choice of $\med,\ \mpd$. We see the largest deviations from the SM scalings at transitions to and from the ground state ($1,0$). The largest of these ground state deviations from SM scalings occurs at low temperatures ($T_\gamma = 0.01 \bd$ with $\mathcal{R}_m = 0.1$). In the case of photo emission there are $O(1)$ discrepancies. 
 
For other transitions, deviations from the SM scalings grow at large temperatures, but are always smaller than those of the ground-state transitions. We note that for other choices of $\mathcal{R}_m$, this pattern of radiation-temperature dependence may change. For example, at $\mathcal{R}_m = 1$, the ground state transitions deviate from the SM scalings around the same order of magnitude for all three choices of $T_\gamma \propto \bd$. For all choices of temperature studied, the ground state transitions show the largest deviations compared to all others studied. 

We have shown that the SM scaling for the recombination coefficient $\alpha_{21}$ is accurate within a few percent. Additionally, although bound-bound transitions to the ground state (1s) can deviate from the SM scalings up to $\mathcal{O}(1)$, intermediate transitions deviate near the percent level or less (see~\autoref{fig:converged_percent_difference} for specific transitions). If it is possible to use SM scalings, the bound-bound transition rates are solvable recursively, and one can calculate the  multi-level atom effective recombination coefficient quickly for hundreds of internal states (\cite{Ali-Haimoud:2010} uses $n_{\text{max}}=500$). Beyond the SM assumptions, we have seen that the rates are much more complicated and cannot be solved recursively; therefore, an exact calculation of $\mathcal{A}_{21}$ is computationally expensive. 

Instead, we vary the effective recombination coefficient $\mathcal{A}_{2\ell}$ at the $\mathcal{O}(1)$ level to test the effect on cosmological observables using a modified version of the cosmological Boltzmann code CLASS~\cite{CLASS_overview,Blas:2011rf, Bansal:2022,Bansal:2022qbi}, and find that the CMB power spectrum is affected only at the $\mathcal{O}(10^{-5})$ level. The largest effect was a modification at the $\mathcal{O}(1)$ level of the dark ionization fraction that freezes out, which is not directly observable. Moving forward, we use tabulated effective recombination coefficients, photo-ionization rates, and bound-bound transition rates that use the SM scalings, as used in the recombination code HyRec~\cite{Ali-Haimoud:2010hou,Lee:2020obi}. 

\section{Cosmological Constraints} \label{sec:cosmology} 
We have validated the SM scalings for radiative processes as being sufficiently accurate for predicting the cosmological history surrounding dark recombination with respect to its impact on observables like the CMB. This allows us to confidently explore a wider parameter space compared to what was previously done.  

In addition, the Atacama Cosmology Telescope (ACT) with its Data Release 6 (DR-6) extended measurements of the CMB angular power spectra to higher $\ell$ with lower uncertainties than Planck, particularly in polarization~\cite{ACT:2025fju}. We here derive new leading constraints on aDM from large-scale cosmology, allowing for the first time the dark electron and dark proton to realize the equal-mass, positronium-like limit, and allowing for $\alphad$ as high as 0.3. We note that previous work that calculated limits on aDM using the Planck 2018 data release required $\med \leq \mpd/10$ and $\alphad \leq 0.2$~\cite{Bansal:2022qbi}. However, this previous study had not validated the re-scaled SM rates for this range of aDM parameter space. 

We briefly review the impact of aDM on large-scale cosmological observables before describing the Bayesian Markov Chain Monte Carlo analysis we perform, and the resulting bounds we obtain. Where appropriate, we highlight the ways in which ACT DR-6 data better constrains the aDM model compared to Planck 2018 data.  

\subsection{Large-Scale Cosmology of Atomic Dark Matter}

The atomic dark sector affects the angular power spectrum of the CMB in a complex way determined by the interplay of the effects of additional radiation, its transition from scattering to free-streaming, and the aDM undergoing dark acoustic oscillations (DAO). Additional radiation increases the expansion rate of the universe during radiation domination, leading to increased Silk damping of the high-$\ell$ tail of the CMB~\cite{Hou:2011ec}. However, its scattering nature at early times induces a positive phase-shift to higher $\ell$ relative to free-streaming radiation, which occurs primarily for high $\ell$ that are sensitive to $k$-modes that entered the horizon before the dark sector decoupled. At the same time, the gravitational perturbation is reduced due to aDM undergoing DAOs instead of falling into over-densities, shifting the equilibrium point of photon density perturbations to lower values~\cite{Bansal:2022}. This increases the amplitude of expansion peaks and decreases the amplitude of compression peaks in the power spectrum. The combined effect is a complicated oscillatory modification of the CMB, whose amplitude grows with $\fd$ and is largest at high $\ell$. 

The dark sound horizon at the time the dark radiation and aDM kinetically decouple from each other, $\rdao$, is the key quantity that defines the length scale at which effects of the atomic dark sector begin to appear. At scales smaller than this, the matter power spectrum is suppressed and displays DAOs. In the CMB, $\ell$-modes that are primarily sensitive to scales smaller than $\rdao$ are the most strongly affected.

The $\lcdm$ limit is realized in the aDM parameter space in the limit where $\rdao\rightarrow 0$ or $\fd\rightarrow 0$, and $\deltaN \rightarrow 0$. For $\deltaN > 0$ but $\rdao\rightarrow 0$ or $\fd \rightarrow 0$, the model is effectively $\lcdm$ plus free-streaming dark radiation. Generally, smaller $\deltaN$, larger $\med$, and larger $\alphad$ lead to smaller $\rdao$. This is the case when the exponential decrease in free dark electron number density (due to dark recombination) triggers the decoupling of the dark plasma. In such cases, lowering the temperature of the dark sector causes the dark sector to undergo dark recombination at earlier times. Raising $\med$ or $\alphad$ raises the binding energy of the dark hydrogen, thereby increasing the temperature at which dark recombination occurs. In the hydrogen-like limit of $\mpd \gg \med$, the dark binding energy does not depend on $\mpd$, and $\rdao$ is very weakly dependent on $\mpd$. In the positronium-like limit, the dark proton and dark electron mass have equal influence on the dark hydrogen binding energy and $\rdao$.

If the dark sector is sufficiently weakly coupled and has low enough number density that the dark radiation and aDM decouple before dark recombination, the correlation of $\rdao$ with $\alphad$ reverses. Because $\sigma_{T}\propto \alphad^{2}/\med^{2}$, lower $\alphad$ leads to a lower Thomson scattering rate, earlier decoupling, and lower $\rdao$. In the parameter scans we perform, this limit is only realized for relatively large dark proton mass, $\mpd=1\ \TeV$, since the dark atomic number density is $\propto 1/m_{H_{D}}$ for fixed $\fd$. 

\subsection{Datasets}

We obtain constraints for two different sets of observations, in order to demonstrate the difference in constraining power between Planck and ACT. The first is the combination of datasets referred to as `P-ACT-LBS' used to study extended models in~\cite{ACT:2025tim}, which comprises the following datasets:
\begin{itemize}
    \item CMB TTTEEE power spectra from ACT DR6~\cite{ACT:2025fju}.  
    \item CMB TTEEEE power spectra from Planck 2018, including only $\ell < 1000$ in TT and $\ell < 600$ in TE/EE~\cite{Planck:2019nip}.
    \item Combined CMB lensing power spectrum from ACT DR6 and Planck PR4~\cite{Carron:2022eyg,ACT:2023dou,ACT:2023kun}.
    \item Measurements of the BAO feature from DESI Year-1 using galaxy, quasar, and Lyman-$\alpha$ forest tracers spanning redshifts $0.1 < z < 4.2$~\cite{DESI:2024lzq,DESI:2024mwx,DESI:2024uvr}. 
    \item Type Ia supernovae from the Pantheon+ catalogue~\cite{Scolnic:2021amr,Brout:2022vxf}.
\end{itemize}

The second combination of datasets does not include any power spectra from ACT DR6 and instead includes the TTTEEE power spectra from Planck without cutting on $\ell$ as well as the lensing power spectrum only from Planck. 
We denote this combination of datasets as `P-LBS'. Each of these datasets have publicly available likelihoods implemented in the cosmological sampling code $\texttt{cobaya}$, which we use to perform Markov Chain Monte Carlo scans of the model parameter space~\cite{Lewis:2002ah,Lewis:2013hha,Torrado:2020dgo}. We use the foreground-marginalized $\texttt{ACT-lite}$ and $\texttt{plik\_lite}$ likelihoods. This code interfaces with a version of the cosmological Boltzmann solver $\texttt{CLASS}$~\cite{CLASS_overview,Blas:2011rf}, modified to include an atomic dark sector in order to compute the necessary cosmological observables~\cite{Bansal:2022,Bansal:2022qbi}. Furthermore, we have adjusted the \texttt{CLASS-aDM} code to correctly treat the positronium limit (in particular, replacing the electron mass with the reduced mass when appropriate) and made improvements to the stability of the code. Our scans use the same precision settings in $\texttt{CLASS}$ as specified in~\cite{ACT:2025tim}. 

\subsection{Scan strategy}
Because of the many limits that reduce to $\lcdm$ in the five-dimensional aDM parameter space, $\rdao$ is the quantity we can most robustly constrain along with $\deltaN$ using the CMB. However, we cannot sample it directly, because $\rdao$ is a derived function of all five aDM parameters that must be computed numerically with $\texttt{CLASS}$. In order to obtain informative constraints, we perform several scans allowing three of the aDM parameters to vary along with the $\lcdm$ parameters. To ensure that the $\lcdm$ limit is realized in our scans, we allow $\deltaN$ to vary. To show how the cosmology constraints behave in the positronium-like limit as well as for large $\alphad$, we allow the dark electron mass $\med$ and dark fine structure constant $\alphad$ to vary, while holding the dark proton mass $\mpd$ fixed. The bound on $\rdao$ is strongly dependent on $\fd$, so we fix it in each of our scans. 

We perform scans for $\mpd \in \{1\ \MeV,\ 1\ \GeV,\ 1\ \TeV\}$. For each dark proton mass, we fix $\fd \in \{0.05,\ 1\}$, and for $\mpd=1\ \GeV$ we additionally run scans at $\fd =0.01,\ 0.1$, for a total of eight different aDM scans. We employ uniform priors on $\deltaN$, $\log_{10}(\med/\GeV)$, and $\alphad$. For each choice of $\mpd$, the upper bound of $\med$ realizes the positronium-like limit. The prior ranges are given in~\autoref{tab:priors}. 

\begin{table} [h]
        \centering
        \begin{tabular}{ c c c }
        \hline
        Parameter & Minimum & Maximum \\
        \hline
        $\deltaN$ & 0.0001 & 1 \\
        $\log_{10}(\frac{\med}{\GeV})$ & -4.7 & $\log_{10}(\frac{\mpd}{\GeV})$ \\
        $\alphad$ & 0.005 & 0.3 \\
        \hline
        \end{tabular}
        \caption{Prior ranges for aDM parameters in our MCMC scans.}
        \label{tab:priors}
    \end{table}

The prior on $\rdao$ is not uniform, and its bounds vary with each of the sampled aDM parameters. When viewing 95\% confidence level contours in 2D-marginalized slices of parameter space including $\rdao$ or considering one-dimensional bounds on $\rdao$, care must be taken that these reflect real preferences in the data, and are not prior-dominated. One obvious example of this effect is that while $\rdao \rightarrow 0$ recovers the $\lcdm$ limit, which is known to be a good fit to the data, the bounds on the $\med$, $\alphad$, and $\deltaN$ priors impose a lower bound on $\rdao$. Similarly, there is a prior-induced upper bound on $\rdao$, which may in principle overlap with the region of parameter space that is allowed by data, creating an artificial upper bound on the $\rdao$ posterior. We test for this by computing the 95\%~C.L.~limits on the $\rdao$ prior as a function of each of the sampled aDM parameters, marginalizing over the other two. If this~95\% C.L.~limit overlaps with the 95\%~C.L.~region of the posterior, then the limit is likely prior-dominated. From our scans, we find that this only occurs for $\fd=0.01$, $\mpd=1\ \GeV$.  

The MCMC scans are run using the Metropolis-Hastings method. For all scans that only use Planck CMB data, the MCMC chains are terminated once achieving a Gelman-Rubin convergence criterion of $R-1 \leq 0.01$. Due to the computational limitations of running the much slower ACT DR6 likelihood, scans that include ACT data have a looser convergence criterion, $R-1 \leq 0.05$. Higher convergence is reached in some cases. For example, for $\mpd = 1$ GeV and $f_\chi =0.05$ the scan achieves $R-1 \leq 0.02$. 

\begin{figure}[ht!]
    \centering
    \includegraphics[width=0.95\linewidth]{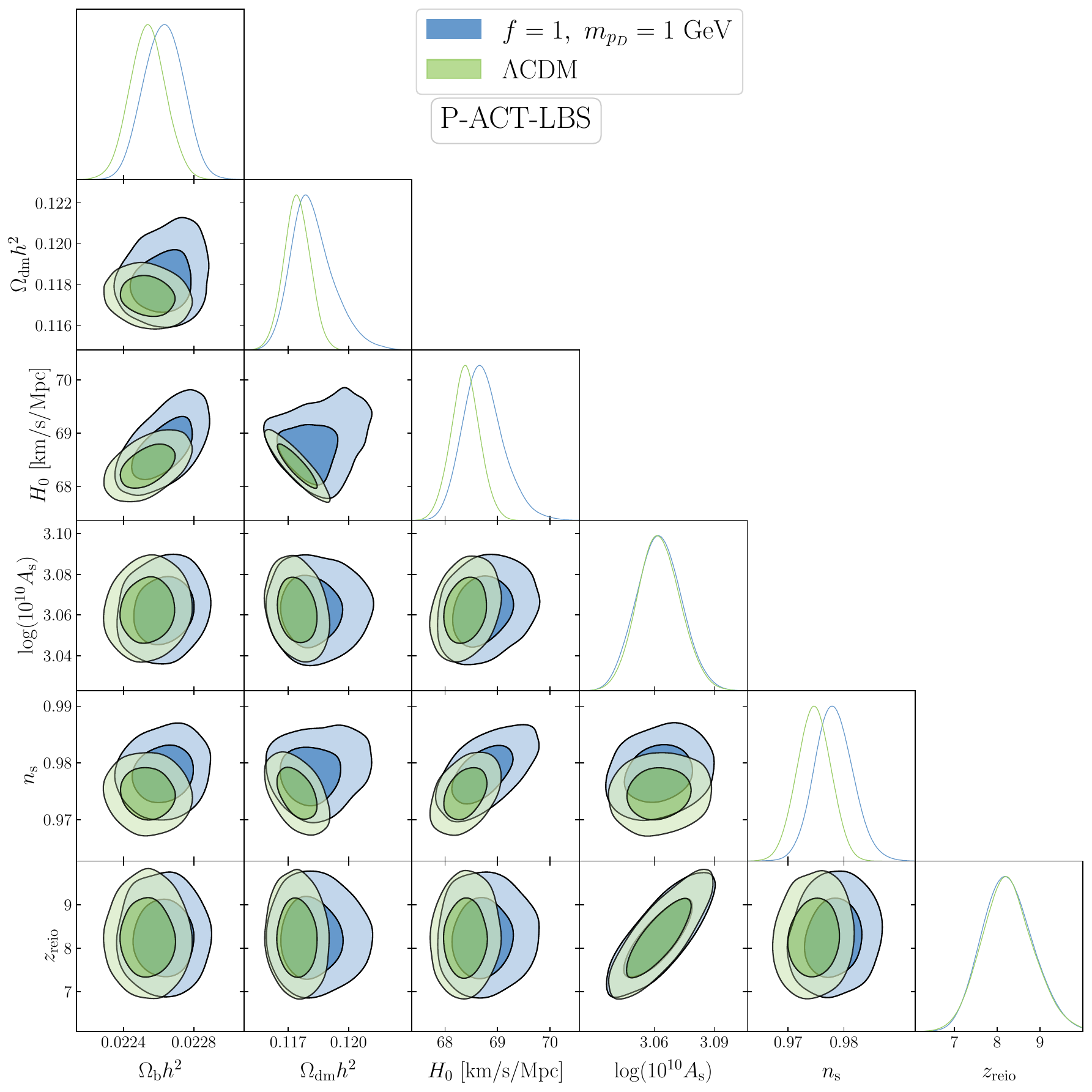}
    \caption{Constraints at 68\% and 95\% C.L.~on the $\lcdm$ parameters from the P-ACT-LBS dataset combination for both CDM (green) and aDM (blue) with $\fd=1$, $\mpd=1$ GeV. The preferred ranges of $H_{0}$ and $\Omega_{\mathrm{dm}}h^{2}$ extend to higher values than allowed in CDM, and the preferred values of $n_{s}$ and $\Omega_{b}h^{2}$ are shifted to slightly higher values. }
    \label{fig:LambdaCDM_constraints}
\end{figure}
\clearpage

\subsection{Results}

We first discuss the inferred $\lcdm$ parameters in our scans, comparing results between CDM and aDM, and between the P-LBS and P-ACT-LBS datasets. The best-fit $\lcdm$ parameters in all of our aDM scans are compatible with the preferred values in CDM, with some notable differences.  First, the addition of dark radiation allows the Hubble constant to range to higher values. The dark matter and baryon abundances $\Omega_{\mathrm{dm}}h^{2}$ and $\Omega_{\mathrm{b}}h^{2}$, as well as the scalar index $n_{s}$, also shift to higher values, although still consistent at the $1\textrm{-}\sigma$ level with CDM results. 
\autoref{fig:LambdaCDM_constraints} displays the 2D marginalized preferred parameter regions for CDM and aDM with $\fd=1$, $\mpd=1\ \GeV$ using the P-ACT-LBS datasets. The 95\% C.L. contours are similar for the other choices of $\fd$ and $\mpd$.

\begin{figure}[t!]
    \centering
    \includegraphics[width=0.50\linewidth]{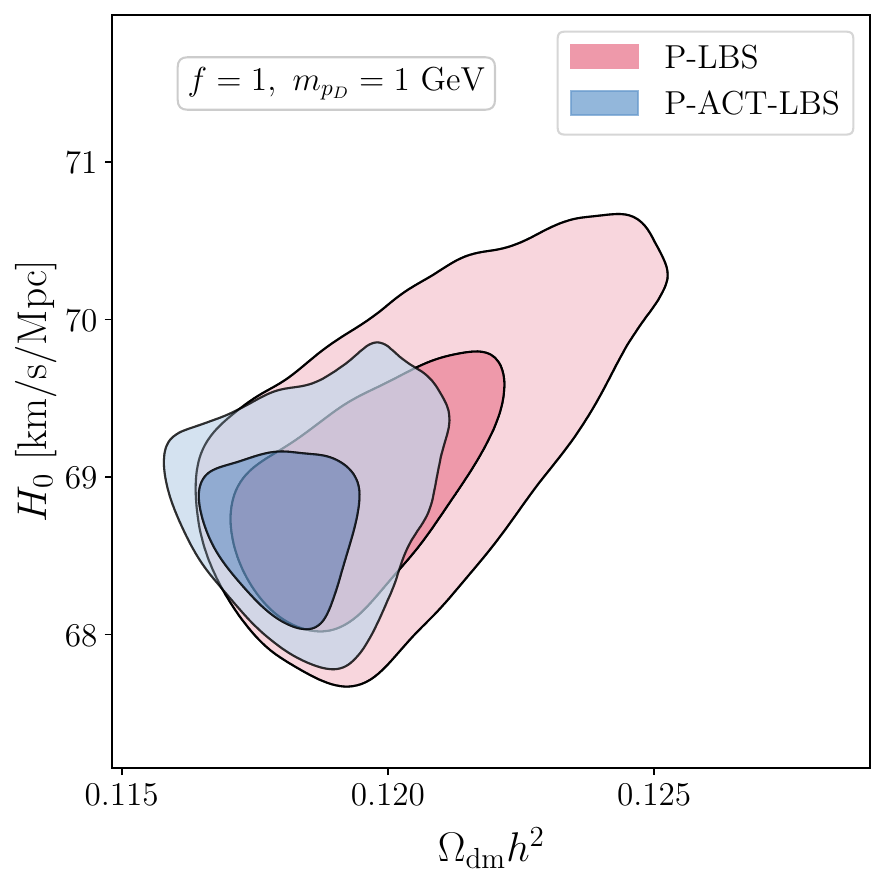}
    \caption{Constraints at 68\% and 95\% C.L. on $H_{0}$ and $\Omega_{\mathrm{dm}}h^{2}$ for aDM with $\fd=1\ \GeV$ and $\mpd=1\ \GeV$ from both P-LBS and P-ACT-LBS dataset combinations. Larger $H_{0}$ and $\Omega_{\mathrm{dm}}h^{2}$ are allowed by Planck than ACT.}
    \label{fig:H0OmegaDM_constraints}
\end{figure}

Within $\lcdm$, the inclusion of ACT data shifts best-fit values and reduces uncertainties relative to Planck~\cite{Planck:2018vyg,ACT:2025fju}. The same occurs for aDM. The most illustrative example of this is the joint constraint on $H_{0}$ and $\Omega_{\mathrm{dm}}h^{2}$. The P-LBS dataset including only Planck CMB data allows the 95\% C.L.~region to extend to higher values in both of these parameters than the dataset including ACT results, as shown in~\autoref{fig:H0OmegaDM_constraints}. This is consistent with ACT's tighter constraint on $\Delta N_{\mathrm{eff}}$ compared to Planck. 

The data exhibit no significant preference for non-zero $\deltaN$ or $\rdao$ in any of our scans, indicating that neither the extra radiation nor the dark matter-dark radiation interactions present in aDM improve the agreement with data. The largest difference in best-fit $\chi^{2}$ values relative to CDM among our scans was $\Delta\chi^{2}=-1.3$ for the P-LBS dataset, and $\Delta\chi^{2}=-2.7$ for the P-ACT-LBS dataset. With three new model parameters, this is not a significant improvement in the fit. 

\begin{figure}[t!]
    \centering
    \includegraphics[width=0.75\linewidth]{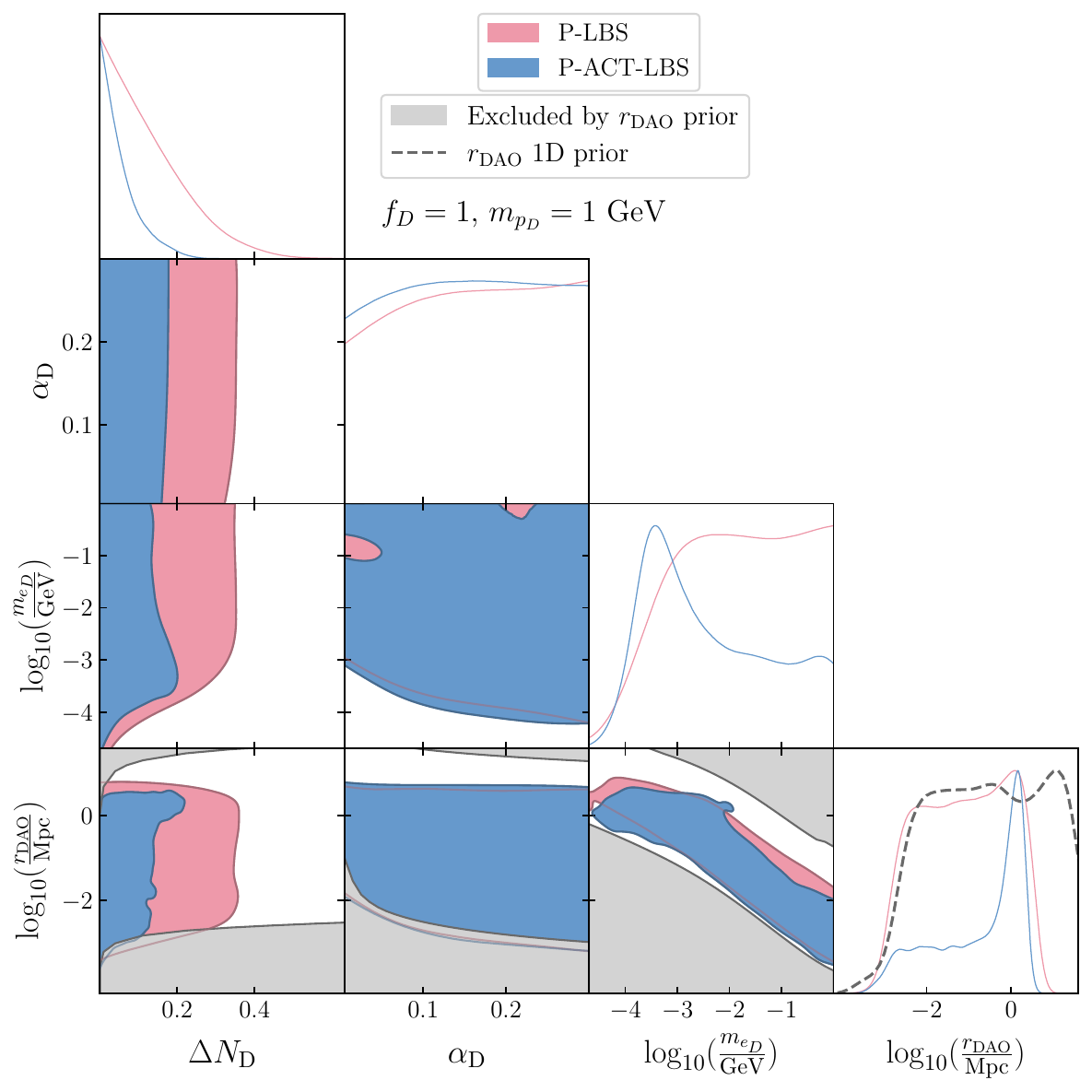}
    \caption{Constraints on dark electron mass $\med$, dark fine structure constant $\alphad$, additional radiation $\deltaN$, and dark sound horizon $\rdao$ at 95\% confidence level, with dark matter fraction $\fd=1$ and dark proton mass $\mpd=1\ \GeV$. Constraints derived using both Planck and ACT CMB data are shown in blue, while constraints using only Planck data are shown in red. $\deltaN$ is constrained to smaller values with the inclusion of ACT data. 
    The effective $\rdao$ prior is given by the distribution of $\rdao$ values resulting from uniformly sampled $\deltaN$, $\alphad$, and $\log_{10}(\frac{\med}{\GeV})$. Values of $\rdao$ outside the central 95\% interval of the derived $\rdao$ prior are shaded gray. In the bottom right plot, the effective one-dimensional $\rdao$ prior is shown in dashed gray. The apparent lower bound on $\rdao$ results from its effective prior going to zero, while the upper bound reflects a preference in the data. }
    \label{fig:3D_f1_mp1GeV}
\end{figure}

We now discuss the bounds we derive on the aDM parameter space. \autoref{fig:3D_f1_mp1GeV} shows 2D marginalized 95\% C.L. regions for the sampled aDM parameters as well as $\rdao$, for both the P-LBS and P-ACT-LBS datasets. The aDM fraction is fixed to $\fd=1$ and the dark proton mass is $\mpd=1\ \GeV$. Additionally, we show the parameter space `excluded' by the $\rdao$ prior at 95\% shaded in gray, as well as the 1D marginalized $\rdao$ prior overlaid with the $\rdao$ posterior. Both dataset combinations give upper bounds on $\deltaN$ and $\rdao$. The apparent lower bound on $\rdao$ is entirely prior-driven, while the upper bound is far from the edge of the $\rdao$ prior, indicating that it is primarily driven by variation in the likelihood. Simultaneously small values of $\alphad$ and $\med$, corresponding to low dark hydrogen binding energy, are disfavored. 

For $\med \gtrsim 1\ \MeV$, the bound on $\deltaN$ saturates to the constraint on additional free-streaming radiation, $\deltaN < 0.35$ for Planck and $\deltaN < 0.16$ for ACT. At lower values of $\med$, $\deltaN$ is constrained to be much smaller in order to achieve a small enough $\rdao$ to agree with the data. 

With the addition of the new ACT-DR6 data we find much stronger bounds on $\deltaN$ as well as a mildly stronger bound on $\rdao$ compared to the constraints using the P-LBS dataset. The qualitative features of the constraints we have described are common to all the scans we performed, from $\fd=0.01$ to $1$ and $\mpd=1\ \MeV$ to $1\ \TeV$. Because the most robust bounds are on $\rdao$ and $\deltaN$, we show further results of our other scans in the space of these two parameters.  

\begin{figure}[t!]
    \centering
    \includegraphics[width=0.50\linewidth]{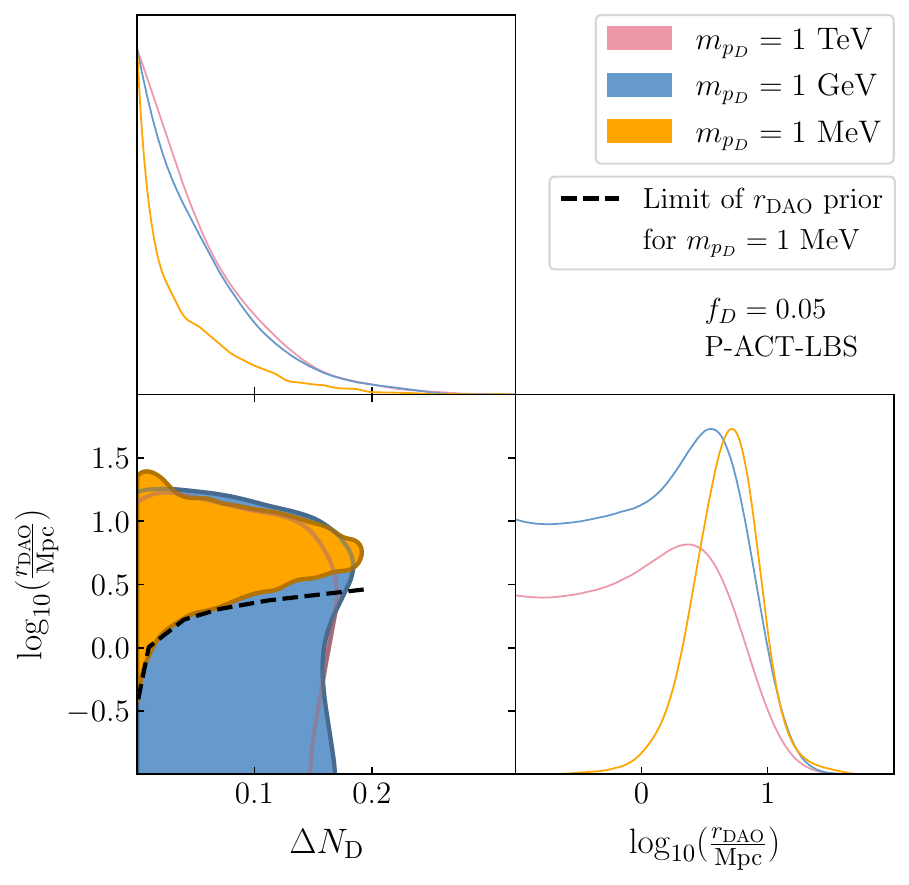}
    \caption{Constraints on $\rdao$ and $\deltaN$ at 95\% confidence level for three choices of the dark proton mass with $\fd=0.05$, using ACT DR6 CMB data, the Pantheon+ supernova catalogue, and BAO measurements from DESI 2024 (P-ACT-LBS). As discussed in~\autoref{fig:3D_f1_mp1GeV}, there is a lower bound on $\rdao$, which is prior-driven and unphysical. For the $y$-range shown in this plot, this is only visible for $\mpd=1$ MeV, and is marked by a dashed black line. $\deltaN$ is restricted to be $\lesssim 0.16$ and $\rdao$ has an upper bound $\lesssim 10\ $Mpc for all three choices of $\mpd$.}
    \label{fig:3D_f0.05}
\end{figure}

\begin{figure}[t!]
    \centering
    \includegraphics[width=0.50\linewidth]{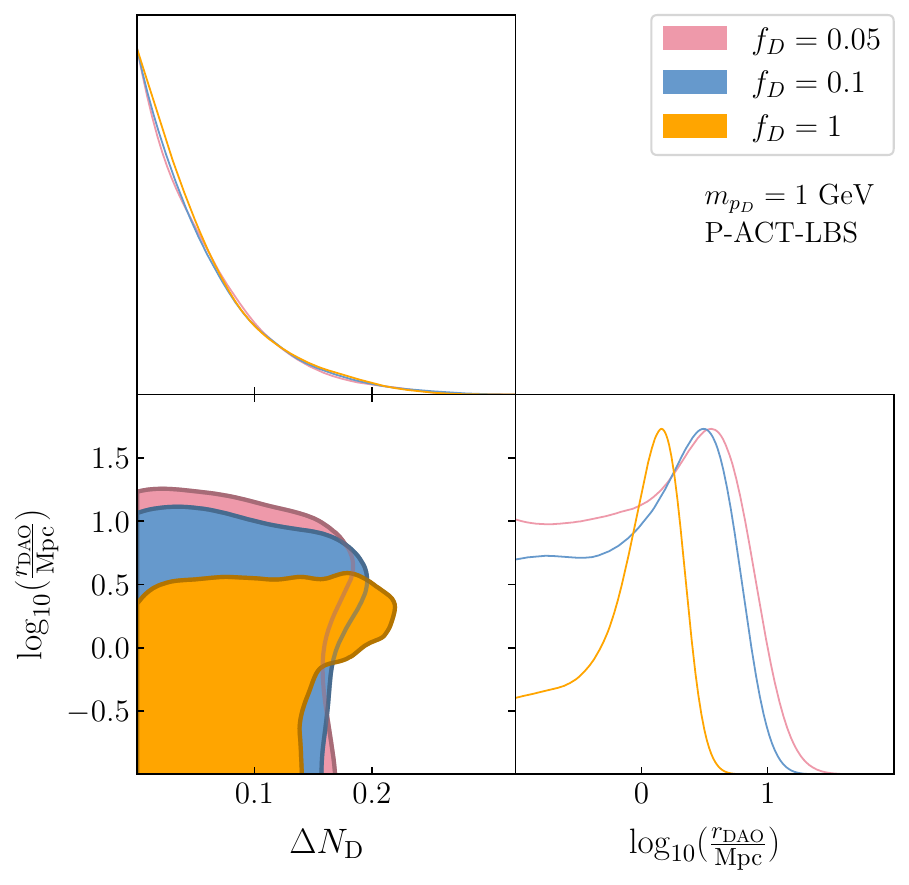}
    \caption{Constraints on $\rdao$ and $\deltaN$ at 95\% confidence level for three choices of $\fd$ with $\mpd=1\ \GeV$, using ACT DR6 CMB data, the Pantheon+ supernova catalogue, and BAO measurements from DESI 2024 (P-ACT-LBS). There is an upper bound on $\rdao$, which increases for smaller $\fd$. $\deltaN$ is constrained to be $\lesssim 0.2$ for all three values of $\fd$. A similar scan was performed for $\fd=0.01$. The upper bound on $\rdao$ was found to be prior-dominated, so we omit it from this plot.}
    \label{fig:3D_mp1GeV}
\end{figure}

The upper bound on $\rdao$ is strongly dependent on the fraction of dark matter $\fd$ that is atomic, and has almost no dependence on $\mpd$. In~\autoref{fig:3D_f0.05}, we show the 2D constraints on $\deltaN$ and $\rdao$ for scans with $\fd=0.05$ and three different values of $\mpd$. While the lowest $\rdao$ allowed at 95\% confidence varies strongly with $\mpd$, which sets the upper bound of the $\med$ prior, the upper boundary of the confidence region has no such dependence on $\mpd$. The 95\% confidence level upper bound on $\rdao$ for $\fd=0.05$ ranges from 7.6 Mpc for $\mpd=1\ \TeV$ to 12 Mpc for $\mpd=1\ \MeV$. In~\autoref{fig:3D_mp1GeV}, we show the constraints for fixed $\mpd=1$ GeV and varying $\fd$. The upper bound on $\rdao$ decreases as $\fd$ increases, as we expect.  

\clearpage

\begin{figure} [t!]
    \centering
    \includegraphics[width=0.80\linewidth]{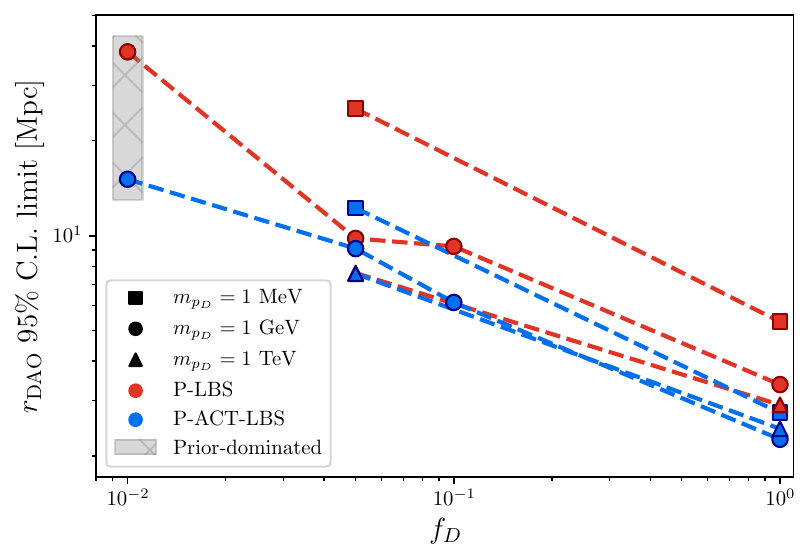}
    \caption{Upper bound on $\rdao$ at 95\% confidence level for each of our scans. We show the constraints for the P-LBS dataset in red and the P-ACT-LBS dataset in blue. We include here the bounds for $\mpd = 1~\textrm{GeV}, \fd = 0.01$, which are consistent with our other results, but emphasize that the $\rdao$ posterior becomes prior-dominated at such small $\fd$.}
    \label{fig:r_DAO}
\end{figure}

We summarize the upper bounds on $\rdao$ from all of our scans in~\autoref{fig:r_DAO}. As expected, the ACT data place a tighter constraint on the DAO scale by up to a factor of two compared to Planck. The limits we find using Planck data are consistent with those found in~\cite{Bansal:2022qbi}. Larger aDM fractions are more tightly constrained, with the bound on $\rdao$ scaling approximately as~$f_{D}^{-2/5}$. The bound is fairly insensitive to $\mpd$, varying by at most a factor of 1.6 across six orders of magnitude in $\mpd$ when using the P-ACT-LBS dataset. 

The bounds on $\deltaN$ are for the most part independent of $\fd$ and $\mpd$. The only exception is $\fd=1$, $\mpd=1\ \MeV$, where the bound on $\deltaN$ is significantly tighter. This is because when the dark proton mass (and therefore also dark electron mass) is low, the dark sector temperature is required to be low in order to yield small enough $\rdao$ to agree with data. We list the 95\%~C.L.~one-dimensional bounds on $\rdao$ and $\deltaN$ in~\autoref{tab:constraints}.

\begin{table}
\centering
\begin{tabular} { l l c c}
$\fd$ & $\frac{\mpd}{\GeV}$& $\frac{\rdao}{\mathrm{Mpc}}$ 95\% limit & $\deltaN$ 95\% limit\\
\hline
0.05 & 0.001 & 12& 0.14 \\
1 & 0.001 & 2.7 & 0.087\\
0.01 & 1 & prior-dominated& 0.16\\
0.05 & 1 & 9.1& 0.16\\
0.1 & 1 & 6.1& 0.16\\
1 & 1 & 2.3 & 0.16\\
0.05 & 1000 & 7.6& 0.15\\
1 & 1000 & 2.4& 0.15\\
\hline
\end{tabular}
\caption{Marginalized constraints on $\rdao$ and $\deltaN$ using Planck, ACT, BAO, and Pantheon+ (P-ACT-LBS) data.}
\label{tab:constraints}
\end{table}

\clearpage
\section{Conclusions} \label{sec:Conclusion}
Through its close but inexact mapping to the SM hydrogen system, atomic dark matter offers us the opportunity to re-examine the familiar from a new perspective. Freed from the usual assumptions that the proton is effectively stationary and that the fine structure constant is extremely small, we have calculated radiative transition rates and cross-sections for dark hydrogen in generality from first principles. The power-law parameter dependencies derived using the SM approximations are sufficient to predict the rates that dominate dark hydrogen recombination with at worst $\mathcal{O}(0.1)$ error when compared to full calculations, for dark electron masses from the hydrogen-like to positronium-like limits and dark fine structure constants up to $0.3$. Bound-bound transitions to the ground state, which do not contribute directly to the case-B recombination rate, can have $\mathcal{O}(1)$ error. At higher values of $\alphad$, fine structure and relativistic corrections become comparable to the classical hydrogen energy levels. We find that most of the difference between re-scaled SM radiative bound-bound transition rates and our full calculation can be accounted for by correctly tracking momentum transfer to the center-of-mass degree of freedom, and provide an easily calculable correction factor that greatly improves agreement with our full calculation. 

While our conclusion that re-scaled SM rates are sufficiently accurate for radiative transitions well beyond the SM-like regime cannot be directly transferred to collisional or molecular processes, it suggests that the same might be true in general. We leave an investigation of these more complex processes to future work. 

Ignorance of the limits on the range of validity of rescaled SM rates has prevented previous work investigating the cosmological phenomenology of atomic dark matter from fully exploring the model parameter space. We have verified for the first time that re-scaled SM rates for the processes involved in recombination of the atomic dark sector are sufficiently accurate to use when computing large-scale cosmological observables like the CMB within atomic dark matter. 
Having validated these re-scaled rates, we take advantage of new measurements of the CMB from the ACT collaboration to obtain constraints on atomic dark matter over an expanded parameter space compared to previous work. Our results are the most stringent limits yet on atomic dark matter from large-scale cosmology. We find that the contribution to $N_{\mathrm{eff}}$ due to dark photons must be $\deltaN \leq 0.16$ at 95\% confidence level, significantly improving on the Planck constraint. The dark sound horizon scale $\rdao$ is constrained to be $\lesssim 10$~Mpc for atomic dark matter fraction $\fd = 0.05$, and $\lesssim 2.5$~Mpc for $\fd=1$.

\acknowledgments
We thank Yacine Ali-Ha\"imoud, Mariangela Lisanti, David Curtin, and Sandip Roy for useful discussions. JB acknowledges support from NSF grants PHY-2210533 and PHY-2513893. RE, MHM, and GS acknowledge support from DOE Grant DE-SC0025309 and Simons Investigator in Physics Awards~623940 and MPS-SIP-00010469. J.P.-R. acknowledges the support of the Simons Foundation. 

\appendix

\section{Transition Amplitudes} \label{sec:overlaps}
We compute four overlaps, $S_{jf}$, where the index $j\in \{1,2\}$ indicates the overlap structure, either~\cref{eqn:overlapS1}, which does not have a momentum operator, or~\cref{eqn:overlapS2}, which contains a 
momentum operator that acts on the initial state, and $f \in \{b, c\}$ indicates whether the final state is a bound or continuum state. In all cases, the initial state is bound. Therefore, the initial state is given by $\ket{nlm}$ and the 
final state is given by $\ket{n_f l_f m_f}$ if bound or $\ket{k_f l_f m_f}$ if continuum. The goal of this appendix is to simplify these four overlaps by analytically solving the relevant integrals that arise. 
We can write the overlap in integral form and Taylor expand the exponential in~\cref{eqn:overlapS1} and~\cref{eqn:overlapS2} to find,
\begin{align}\label{eqn:overlapSij}
    S_{jf} = \sum_{\zeta = 0}^{\infty} \left(\frac{i \omega' \mud c \alphad^2}{4 n^2 \hbar}\right)^{\zeta} \frac{h_j(\zeta)}{\zeta!} \int dr d\chi d\phi r^{2+\zeta} \chi^{\zeta} \psi_{f}^{*} \hat{O}_j \psi_{nlm} \ ,
\end{align}
where we have written the angular integral in terms of $\chi \equiv \cos \theta$. Additionally, $\psi_{f}$ represents the final state wavefunction (either bound or continuum) and $\psi_{nlm}$ the initial state wavefunction (bound in all cases) is defined in~\cref{eqn:psinlm}. We define $\hat{O}_{1}=1$ and $\hat{O}_{2}=\frac{d}{dx}$. The auxiliary function $h_j (\zeta)$ depends only on aDM masses and $\zeta$, with $h_1(\zeta) = (-1)^{\zeta} \left[1+\sqrt{1-y}\right]^{\zeta} - \left[1-\sqrt{1-y}\right]^{\zeta}$ and $h_2(\zeta) = \frac{2 i \hbar}{y \mHd} h_1(\zeta+1)$. Note that the dipole approximation would keep only the $\zeta=0$ term of the expansion. 

Since $l + 1 - n \in \mathbb{Z} \leq 0$ for bound states, the confluent hypergeometric function's Taylor series expansion truncates, allowing \cref{eqn:psinlm} to be simplified to 
\begin{align}
\psi_{nlm}(r, \theta, \phi)   &= N_b \sum_{\gamma = 0}^{n-l-1} \frac{\left(l+1-n\right)^{(\gamma)}}{\left(l+1\right)^{(\gamma)} \gamma!}\left(\frac{1}{a_0 n}\right)^{\gamma} P_{l}^{m}(\chi) e^{i m \phi} e^{-\frac{r}{a_{o} n}} r^{\gamma+l} \ .
\end{align}
Substituting this into our total wavefunction in  \cref{eqn:overlapSij} and making a change of variable $r\rightarrow a_{0}nr$, we obtain 
\begin{align}\label{eqn:Sij}
    &S_{jf} = N_f N_b (a_0 n)^{3+l_f+l-c_j}\sum_{\gamma = 0}^{n-l-1} \frac{\left(l+1-n\right)^{(\gamma)}}{\left(l+1\right)^{(\gamma)} \gamma!}\sum_{\zeta = 0}^{\infty} \left(\frac{i \omega' \alphad}{4 n}\right)^{\zeta} \frac{h_j(\zeta)}{\zeta!} 
    \mathcal{I}_{jf} \ .
\end{align}
We note that $\zeta = 0$ is the dipole term, and higher order terms are subdominant; in practice, we sum to $\zeta=3$. Additionally, we have defined $c_{1}=0$, $c_{2}=1$, and the integral 
\begin{align}
    \mathcal{I}_{jf} \equiv \int dr r^{2+\zeta+l_f} e^{- \frac{n r}{s_f}} F(l_f+1-s_f; 2l_f+2; \frac{2 n r}{s_f}) \int d\chi P_{l_f}^{m_f}(\chi) \chi^{\zeta} \int d\phi   e^{-i m_f \phi}  \hat{O}_j r^{l+\gamma} e^{-r} P_{l}^{m}(\chi) e^{i m \phi} \ .
\end{align}
We define $s_f$ as $s_c = \frac{i n}{k'}$ for a continuum final state or $s_b = n_{f}$ for a bound final state. For convenience, we denote the first integral over $\phi$ as $\mathcal{I}^{\phi}_{jf}$ and the subsequent integral over $\chi$ as $\mathcal{I}^{\chi}_{jf}$. After expanding $\hat{O}_{j}$ in spherical coordinates, we analytically solve $\mathcal{I}^{\phi}_{jf}$ to obtain 
\begin{align}
    \mathcal{I}^{\phi}_{1f} &= 2 \pi \delta^{m_f}_{m} r^{l+\gamma} e^{-r} P_{l}^{m}(\chi) \ ,\\
    \mathcal{I}^{\phi}_{2f} &= \pi \delta_{m}^{m_f \pm 1}\left[\sqrt{1-\chi^2}\frac{d}{dr} - \frac{\chi \sqrt{1-\chi^2}}{r} \frac{d}{d\chi} \pm \frac{m}{r \sqrt{1-\chi^2}}\right] r^{l+\gamma} e^{-r} P_{l}^{m}(\chi) \ .
\end{align}
We then write $\mathcal{I}^{\chi}_{jf} \equiv \int d\chi P_{l_f}^{m_f}(\chi) \chi^{\zeta} \mathcal{I}^{\phi}_{jf}$ in terms of several integrals over $\chi$, which are easily computed analytically once $l_f$, $m_f$, $l$, $m$, $\zeta$, and $\gamma$ are specified: 
\begin{align}
   \mathcal{I}^{\chi}_{1f} &= 2 \pi \delta^{m_f}_{m} r^{l+\gamma} e^{-r} V \ ,\\
   \mathcal{I}^{\chi}_{2f} &= \pi \delta_{m}^{m_f \pm 1}\left(V_2 r^{l+\gamma-1} e^{-r} - V_1 r^{l+\gamma} e^{-r}\right) \ ,
\end{align}
 where we have defined the following helpful auxiliary functions
\begin{align}
V(\zeta) &\equiv \int d\chi P_{l_f}^{m}(\chi) \chi^{\zeta} P_{l}^{m}(\chi)\\
 V_1(\zeta) &\equiv \int d\chi P_{l_f}^{m_f}(\chi) \chi^{\zeta} \sqrt{1-\chi^2}P_{l}^{m}(\chi)\\
 V_2(\zeta, \gamma) &\equiv (l+\gamma)V_1  - \int d\chi P_{l_f}^{m_f}(\chi) \chi^{\zeta} \chi \sqrt{1-\chi^2}\frac{d}{d\chi}P_{l}^{m}(\chi) \\
 & \qquad \pm m \delta_m^{m_f \pm 1} \int d\chi P_{l_f}^{m_f}(\chi) \frac{\chi^{\zeta}}{\sqrt{1-\chi^2}}P_{l}^{m}(\chi) \ . 
\end{align}
Finally, we can perform the integral over $r$ by making use of the identity,
\begin{align}\label{J_int}
    J_{\alpha, \gamma}^{\nu, \lambda} (k) \equiv \int_0^{\infty}dr e^{-\lambda r} r^{\nu} F(\alpha, \gamma, k r) \ ,
\end{align}
assuming $\nu \in \mathbb{N}$ and $\text{Re}(\lambda) > |\text{Re}(k)|$. If $\alpha$ is a non-positive integer, the latter condition can be relaxed to $\text{Re}(\lambda) > 0$. Note, these conditions are satisfied for all of the integrals over $r$ in $\mathcal{I}_{jf}$. Therefore, we get 
\begin{align}
    \mathcal{I}_{1j} &= 2 \pi \delta^{m_f}_{m} V(\zeta)  J_{l_f+1-s_j, 2l_f+2}^{2+\zeta+l_f+l+\gamma,1+n/s_j}\left(\frac{2n}{s_j}\right),\\
    \mathcal{I}_{2j} &= \pi \delta_{m}^{m_f \pm 1} \left\{ V_2(\zeta,\gamma) J_{l_f+1-s_j,2l_f+2}^{1+\zeta+l_f+l+\gamma,1+n/s_j}\left(\frac{2n}{s_j}\right)- V_1(\zeta) J_{l_f+1-s_j, 2l_f+2}^{2+\zeta+l_f+l+\gamma,1+n/s_j}\left(\frac{2n}{s_j}\right)\right\},
\end{align}
which can be trivially plugged into~\cref{eqn:Sij} to solve for any of the four overlaps.

\section{Origin of Deviations from the Standard Model} \label{sec:DeviationsFromSM}

As we have seen from the results in the main text, the SM scalings are accurate enough to use for the purpose of computing cosmological observables, for $\alphad \leq 0.3$ and $\med/\mpd\leq 1$. However, it is instructive to study the origin of the discrepancy between the re-scaled SM results and our calculation, which grows large as $\alphad \rightarrow 1$. We will see that the presence of the COM degree of freedom and its coupling to the interaction with dark photons is the cause of most of the discrepancy. It is also responsible for the apparent asymmetry between the bound-bound photo-absorption and emission cross-sections. For values of $\mathcal{R}_{m} \neq 1$, emission is enhanced while absorption is suppressed. Finally, we derive an easily computable correction factor that accounts for almost all of the discrepancy between re-scaled SM rates and our calculation. 

We start by highlighting a few important facts. The transition rate is symmetric under exchange of initial and final states, as it must be:
\begin{align}
    \Gamma_{\{\mathbf{P}^{i},i\} \to \{\mathbf{P}^{f},f\}} = \Gamma_{\{\mathbf{P}^{f},f\} \to \{\mathbf{P}^{i},i\}}.
\end{align}
However, it is easy to show that the rate is not symmetric under exchange of only initial and final relative states (without exchanging COM states):
\begin{align}
    \Gamma_{\{\mathbf{P}^{i},i\} \to \{\mathbf{P}^{f},f\}} \neq \Gamma_{\{\mathbf{P}^{i},f\} \to \{\mathbf{P}^{f},i\}}.
\end{align}
To obtain the cross section of \cref{eq:sigma_rel}, we integrate over the COM momentum. Therefore, when we switch between initial and final states ($i, f$) for the relative degree of freedom, we are not simultaneously switching initial and final COM degree of freedom states ($\boldsymbol{P}^i, \boldsymbol{P}^f$). Thus, we do not expect the emission and photo-absorption cross-sections (averaged and summed over initial and final COM states) to be symmetric. In the SM limit, when the COM degree of freedom becomes negligible, so too is the asymmetry between absorption and emission, and one can recover the usual detailed balance relations.

When both the COM and relative degrees of freedom can gain or lose momentum and energy through the absorption or emission of a dark photon, transitions between particular bound states no longer pick out unique dark photon energies, $\omega$. Instead, a range of $\omega$ contribute, weighted by an exponential factor that is contained in~\cref{eq:sigma_rel}, and which we rewrite here for convenience:
\begin{align} \label{eqn:envelope_func}
    F(\omega)=\exp\left[-\frac{\mHd }{2 T_m  \omega^2} \left(E_{\text{rel}}^f - E_{\text{rel}}^i + \frac{\omega^2}{2 \mHd} \pm \omega \right)^2\right] \ .
\end{align}

For emission processes where $E_{\text{rel}}^i > E_{\text{rel}}^f$, only the term with $+\omega$ contributes, while for absorption processes only $-\omega$ contributes. Clearly, in the SM limit where $E_{\text{rel}}\sim \alphad^{2}\mud \ll \mHd$, this distribution becomes narrowly peaked at $\omega=|E_{\text{rel}}^f - E_{\text{rel}}^i|$. The quadratic term in $\omega$, which arises due to accounting for the COM momentum, shifts the peak of this function to lower $\omega$ for emission and to higher $\omega$ for absorption. When integrating over the dark photon thermal distribution in~\cref{eqn:bbRate}, higher-energy photons have lower occupation number, leading to a suppression in the total transition rate. The cross section in therefore weighted by, 
\begin{equation}\label{eqn:env_withPlanck}
\mathcal{F}(\omega) \equiv \frac{1}{e^{\frac{\omega}{T_{\gamma}}}-1} F(\omega) \ .
\end{equation}

\begin{figure}[t!]
    \centering
    \includegraphics[width= 0.90\linewidth]
    {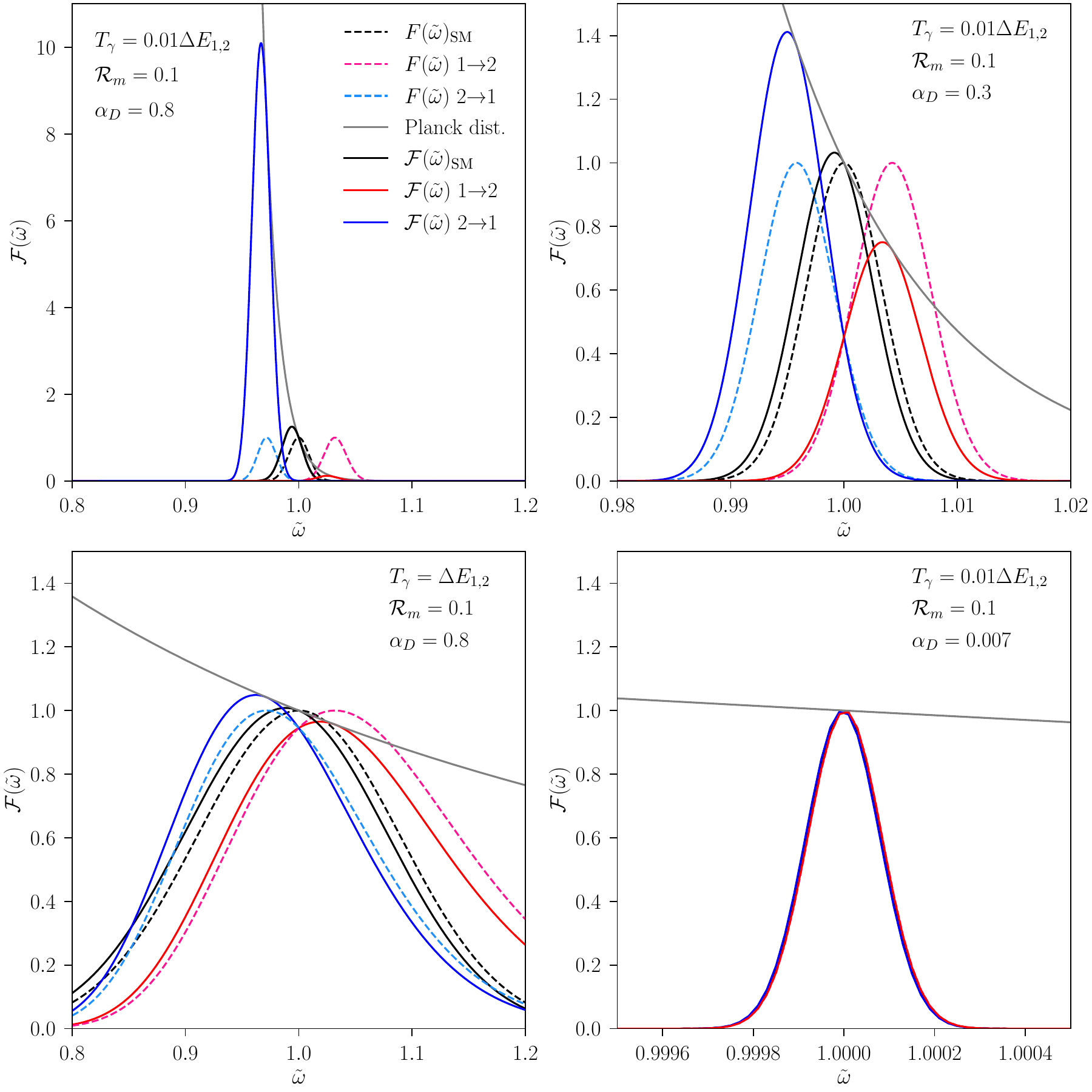}
    \caption{\label{fig:envelopes} We plot the envelope function from~\cref{eqn:envelope_func} and~\cref{eqn:env_withPlanck} for an example photo-absorption $1 \rightarrow 2$ process in dashed pink (solid red) and photo-emission $2 \rightarrow 1$ process in dashed light-blue (solid blue). We compare each to the approximated envelope that ignores the term proportional to $\omega^4$, which accounts for energy transferred to the COM degree of freedom, shown in black. In all cases, $\med = \mpd$ and the matter-ration temperature ratio $\mathcal{R}_m = 0.1$. We plot with respect to a transformed photon energy $\tilde{\omega} \equiv \omega / |E_n - E_{n_f}|$ such that the SM curve is centered at $\tilde{\omega}=1$ and all curves are normalized by $(e^{1/T_{\gamma}}-1)^{-1}$. Note, $\Delta E_{1,2} \equiv E_1 - E_2$. \textbf{Top left:} Smaller deviations occur for smaller $\alpha_D$, and we show the benchmark $\alpha_D=0.3$, which is used throughout the main body of this work. \textbf{Top right:} The largest deviations between the approximated envelope using SM assumptions and the full envelope appear at small radiation temperatures and large $\alpha_D$. \textbf{Bottom left:} Although deviations are large for $\alpha_D = 0.8$, we see the effect of increasing radiation temperature; the photon distribution is much flatter and deviations from the SM estimate decrease. \textbf{Bottom right:} For small $\alpha_D$, the SM estimate is accurate at all temperatures. }
\end{figure}

The behavior of these functions $F$ and $\mathcal{F}$, and their dependence on both $\alphad$ and $T_{\gamma}$, are illustrated in~\autoref{fig:envelopes}. The opposite shifts of $F(\omega)$ to lower and higher $\omega$ for emission and absorption processes, respectively, are displayed in blue and red, while the SM-like limit with negligible quadratic term is shown in black. The effect of weighting by the dark photon distribution is shown by plotting $\mathcal{F}(\omega)$. Energies are normalized to $\tilde{\omega}=\omega/\Delta E_{2,1}$. The left panels of the plot show the dependence of the asymmetry between emission and absorption on $T_{\gamma}$. The Planck distribution is much steeper for lower $T_{\gamma}$, creating a larger asymmetry in the amplitude of $\mathcal{F}(\tilde{\omega})$ between emission and absorption, and larger deviations from the SM value. The right panels of the plot show the dependence on $\alphad$. As $\alphad$ becomes smaller, the offsets of $F(\tilde{\omega})$ from $\tilde{\omega}=1$ decrease, as does the width of $F(\tilde{\omega})$. For $\alphad=0.007$, the full $\mathcal{F}(\tilde{\omega})$ are almost indistinguishable from the SM limit, while at $\alphad=0.007$ deviations in $\mathcal{F}(\tilde{\omega})$ at the $\mathcal{O}(1)$ level start to appear. 

This is the origin of the apparent asymmetry between absorption and emission. Here, we can find an intuitive explanation for this asymmetry. On average, the momentum transfer of emitting or absorbing a photon tends to increase the energy of the COM for both absorption and emission processes. For absorption, since the COM tends to absorb energy, larger photon energies are required to induce the transition. These higher-energy photons are less common, resulting in an overall suppression in the rate. For stimulated emission, the stimulating photon has the same energy as the emitted photon. By Planck's law there are more of these lower-energy photons present, resulting in an overall enhancement in the rate. 

We can account for these effects in a given bound-bound transition rate $R_{nl \rightarrow n_{f}l_{f}}$ by multiplying the usual re-scaled SM rate by a factor $A \equiv \int_{0}^{\infty}d\omega \mathcal{F}(\omega) / \int_{0}^{\infty}d\omega \mathcal{F}(\omega)_\SM$. With this correction, we match $R_{n l \rightarrow n_{f} l_{f}}$ from our full calculation at the $\mathcal{O}(10\%)$ level for $\alphad$ up to 0.8, whereas the usual rescaled SM rates deviate by up to $\mathcal{O}(10^{2})$.

\bibliographystyle{JHEP}
\bibliography{references}

@article{Shandera:2018xkn,
    author = "Shandera, Sarah and Jeong, Donghui and Gebhardt, Henry S. Grasshorn",
    title = "{Gravitational Waves from Binary Mergers of Subsolar Mass Dark Black Holes}",
    eprint = "1802.08206",
    archivePrefix = "arXiv",
    primaryClass = "astro-ph.CO",
    reportNumber = "IGC-18-2-1",
    doi = "10.1103/PhysRevLett.120.241102",
    journal = "Phys. Rev. Lett.",
    volume = "120",
    number = "24",
    pages = "241102",
    year = "2018"
}

@article{Singh:2020wiq,
    author = "Singh, Divya and Ryan, Michael and Magee, Ryan and Akhter, Towsifa and Shandera, Sarah and Jeong, Donghui and Hanna, Chad",
    title = "{Gravitational-wave limit on the Chandrasekhar mass of dark matter}",
    eprint = "2009.05209",
    archivePrefix = "arXiv",
    primaryClass = "astro-ph.CO",
    reportNumber = "LIGO-P2000291",
    doi = "10.1103/PhysRevD.104.044015",
    journal = "Phys. Rev. D",
    volume = "104",
    number = "4",
    pages = "044015",
    year = "2021"
}

@article{Bansal:2022,
  author        = {Bansal, Saurabh and Kim, Jeong Han and Kolda, Christopher and Low, Matthew and Tsai, Yuhsin},
  title         = {Mirror Twin Higgs Cosmology: Constraints and a Possible Resolution to the $H_0$ and $S_8$ Tensions},
  journal       = {Journal of High Energy Physics},
  volume        = {2022},
  number        = {5},
  pages         = {50},
  year          = {2022},
  doi           = {10.1007/JHEP05(2022)050},
  eprint        = {2110.04317},
  archivePrefix = {arXiv},
  primaryClass  = {astro-ph.CO}
}

@article{Peebles:1968ja,
    author = "Peebles, P. J. E.",
    title = "{Recombination of the Primeval Plasma}",
    doi = "10.1086/149628",
    journal = "Astrophys. J.",
    volume = "153",
    pages = "1",
    year = "1968"
}

@ARTICLE{1969JETP...28..146Z,
       author = {{Zel'dovich}, Ya. B. and {Kurt}, V.~G. and {Syunyaev}, R.~A.},
        title = "{Recombination of Hydrogen in the Hot Model of the Universe}",
      journal = {Soviet Journal of Experimental and Theoretical Physics},
         year = 1969,
        month = jan,
       volume = {28},
        pages = {146},
       adsurl = {https://ui.adsabs.harvard.edu/abs/1969JETP...28..146Z}
}

@article{Ali-Haimoud:2010,
  author        = {Ali-Haïmoud, Yacine and Hirata, Christopher M.},
  title         = {Ultrafast effective multilevel atom method for primordial hydrogen recombination},
  journal       = {Phys. Rev. D},
  volume        = {82},
  number        = {6},
  pages         = {063521},
  year          = {2010},
  doi           = {10.1103/PhysRevD.82.063521},
  eprint        = {1006.1355},
  archivePrefix = {arXiv},
  primaryClass  = {astro-ph.CO}
}

@article{Ali-Haimoud:2010hou,
    author = "Ali-Haimoud, Yacine and Hirata, Christopher M.",
    title = "{HyRec: A fast and highly accurate primordial hydrogen and helium recombination code}",
    eprint = "1011.3758",
    archivePrefix = "arXiv",
    primaryClass = "astro-ph.CO",
    doi = "10.1103/PhysRevD.83.043513",
    journal = "Phys. Rev. D",
    volume = "83",
    pages = "043513",
    year = "2011"
}

@article{Lee:2020obi,
    author = {Lee, Nanoom and Ali-Ha{\"\i}moud, Yacine},
    title = "{HYREC-2: a highly accurate sub-millisecond recombination code}",
    eprint = "2007.14114",
    archivePrefix = "arXiv",
    primaryClass = "astro-ph.CO",
    doi = "10.1103/PhysRevD.102.083517",
    journal = "Phys. Rev. D",
    volume = "102",
    number = "8",
    pages = "083517",
    year = "2020"
}

@article{Cyr-Racine:2013fsa,
    author = "Cyr-Racine, Francis-Yan and de Putter, Roland and Raccanelli, Alvise and Sigurdson, Kris",
    title = "{Constraints on Large-Scale Dark Acoustic Oscillations from Cosmology}",
    eprint = "1310.3278",
    archivePrefix = "arXiv",
    primaryClass = "astro-ph.CO",
    doi = "10.1103/PhysRevD.89.063517",
    journal = "Phys. Rev. D",
    volume = "89",
    number = "6",
    pages = "063517",
    year = "2014"
}

@article{Roy:2024bcu,
    author = "Roy, Sandip and Shen, Xuejian and Barron, Jared and Lisanti, Mariangela and Curtin, David and Murray, Norman and Hopkins, Philip F.",
    title = "{Aggressively Dissipative Dark Dwarfs: The Effects of Atomic Dark Matter on the Inner Densities of Isolated Dwarf Galaxies}",
    eprint = "2408.15317",
    archivePrefix = "arXiv",
    primaryClass = "astro-ph.GA",
    doi = "10.3847/1538-4357/adb02b",
    journal = "Astrophys. J.",
    volume = "982",
    number = "2",
    pages = "175",
    year = "2025"
}

@article{Roy:2023zar,
    author = "Roy, Sandip and Shen, Xuejian and Lisanti, Mariangela and Curtin, David and Murray, Norman and Hopkins, Philip F.",
    title = "{Simulating Atomic Dark Matter in Milky Way Analogs}",
    eprint = "2304.09878",
    archivePrefix = "arXiv",
    primaryClass = "astro-ph.GA",
    doi = "10.3847/2041-8213/ace2c8",
    journal = "Astrophys. J. Lett.",
    volume = "954",
    number = "2",
    pages = "L40",
    year = "2023"
}

@article{Burgess:1960,
  author       = {Burgess, A. and Seaton, M. J.},
  title        = {A General Formula for the Calculation of Atomic Photo-ionization Cross Sections},
  journal      = {Monthly Notices of the Royal Astronomical Society},
  volume       = {120},
  number       = {2},
  pages        = {121--151},
  year         = {1960},
  doi          = {10.1093/mnras/120.2.121},
  eprint       = {https://academic.oup.com/mnras/article-pdf/120/2/121/8074665/mnras120-0121.pdf},
}

@article{Burgess1965,
    author = {Burgess, A.},
    title  = {Tables of hydrogenic photoionization cross-sections and recombination coefficients},
    journal = {Monthly Notices of the Royal Astronomical Society},
    year   = 1965, 
    url    = {https://ui.adsabs.harvard.edu/abs/1965MmRAS..69....1B}
}

@article{Erdas:1985,
  author       = {Erdas, Andrea and Mezzorani, G. and Quarati, Piero and Puddu, Gianluigi},
  title        = {Radiative formation of positronium in a vacuum},
  journal      = {Astronomy and Astrophysics},
  volume       = {144},
  pages        = {295--297},
  year         = {1985}
}

@book{landau1981quantum,
  title={Quantum Mechanics: Non-Relativistic Theory},
  author={Landau, L.D. and Lifshitz, E.M.},
  year={1981},
  edition={3rd},
  publisher={Pergamon Press},
  series={Course of Theoretical Physics},
  volume={3},
  address={Oxford},
  note={See Section 36}
}

@book{sakurai2017modern,
  title={Modern Quantum Mechanics},
  author={Sakurai, J.J. and Napolitano, Jim},
  year={2017},
  edition={2nd},
  publisher={Cambridge University Press},
  address={Cambridge}
}

@article{Seager_2000,
doi = {10.1086/313388},
url = {https://dx.doi.org/10.1086/313388},
year = {2000},
volume = {128},
number = {2},
pages = {407},
author = {Seager, Sara and Sasselov, Dimitar D. and Scott, Douglas},
title = {How Exactly Did the Universe Become Neutral?},
journal = {The Astrophysical Journal Supplement Series}
}

@article{Kaplan:2009de,
    author = "Kaplan, David E. and Krnjaic, Gordan Z. and Rehermann, Keith R. and Wells, Christopher M.",
    title = "{Atomic Dark Matter}",
    eprint = "0909.0753",
    archivePrefix = "arXiv",
    primaryClass = "hep-ph",
    doi = "10.1088/1475-7516/2010/05/021",
    journal = "JCAP",
    volume = "05",
    pages = "021",
    year = "2010"
}

@article{Kaplan:2011yj,
    author = "Kaplan, David E. and Krnjaic, Gordan Z. and Rehermann, Keith R. and Wells, Christopher M.",
    title = "{Dark Atoms: Asymmetry and Direct Detection}",
    eprint = "1105.2073",
    archivePrefix = "arXiv",
    primaryClass = "hep-ph",
    reportNumber = "FERMILAB-PUB-11-207-T",
    doi = "10.1088/1475-7516/2011/10/011",
    journal = "JCAP",
    volume = "10",
    pages = "011",
    year = "2011"
}

@article{Cyr-Racine:2012tfp,
    author = "Cyr-Racine, Francis-Yan and Sigurdson, Kris",
    title = "{Cosmology of atomic dark matter}",
    eprint = "1209.5752",
    archivePrefix = "arXiv",
    primaryClass = "astro-ph.CO",
    doi = "10.1103/PhysRevD.87.103515",
    journal = "Phys. Rev. D",
    volume = "87",
    number = "10",
    pages = "103515",
    year = "2013"
}

@article{Fan:2013yva,
    author = "Fan, JiJi and Katz, Andrey and Randall, Lisa and Reece, Matthew",
    title = "{Double-Disk Dark Matter}",
    eprint = "1303.1521",
    archivePrefix = "arXiv",
    primaryClass = "astro-ph.CO",
    doi = "10.1016/j.dark.2013.07.001",
    journal = "Phys. Dark Univ.",
    volume = "2",
    pages = "139--156",
    year = "2013"
}

@article{Foot:2013vna,
    author = "Foot, R.",
    title = "{Galactic structure explained with dissipative mirror dark matter}",
    eprint = "1304.4717",
    archivePrefix = "arXiv",
    primaryClass = "astro-ph.CO",
    doi = "10.1103/PhysRevD.88.023520",
    journal = "Phys. Rev. D",
    volume = "88",
    number = "2",
    pages = "023520",
    year = "2013"
}

@article{Foot:2017dgx,
    author = "Foot, R.",
    title = "{Dissipative dark matter halos: The steady state solution}",
    eprint = "1707.02528",
    archivePrefix = "arXiv",
    primaryClass = "astro-ph.GA",
    doi = "10.1103/PhysRevD.97.043012",
    journal = "Phys. Rev. D",
    volume = "97",
    number = "4",
    pages = "043012",
    year = "2018"
}

@article{Foot:2016wvj,
    author = "Foot, Robert and Vagnozzi, Sunny",
    title = "{Solving the small-scale structure puzzles with dissipative dark matter}",
    eprint = "1602.02467",
    archivePrefix = "arXiv",
    primaryClass = "astro-ph.CO",
    doi = "10.1088/1475-7516/2016/07/013",
    journal = "JCAP",
    volume = "07",
    pages = "013",
    year = "2016"
}

@article{Foot:2015mqa,
    author = "Foot, R.",
    title = "{Dissipative dark matter and the rotation curves of dwarf galaxies}",
    eprint = "1506.01451",
    archivePrefix = "arXiv",
    primaryClass = "astro-ph.GA",
    doi = "10.1088/1475-7516/2016/07/011",
    journal = "JCAP",
    volume = "07",
    pages = "011",
    year = "2016"
}

@article{Fan:2017,
  author       = {Rosenberg, Eliott and Fan, JiJi},
  title        = {Cooling in a dissipative dark sector},
  journal      = {Phys. Rev. D},
  volume       = {96},
  number       = {12},
  year         = {2017},
  doi          = {10.1103/PhysRevD.96.123001},
  eprint       = {1705.10341},
  archivePrefix= {arXiv},
  primaryClass = {astro-ph.CO}
}

@article{Bansal:2022qbi,
    author = "Bansal, Saurabh and Barron, Jared and Curtin, David and Tsai, Yuhsin",
    title = "{Precision cosmological constraints on atomic dark matter}",
    eprint = "2212.02487",
    archivePrefix = "arXiv",
    primaryClass = "hep-ph",
    doi = "10.1007/JHEP10(2023)095",
    journal = "JHEP",
    volume = "10",
    pages = "095",
    year = "2023"
}

@article{Blinov:2021,
    author = "Blinov, Nikita and Krnjaic, Gordan and Li, Shirley Weishi",
    title = "{Toward a realistic model of dark atoms to resolve the Hubble tension}",
    eprint = "2108.11386",
    archivePrefix = "arXiv",
    primaryClass = "hep-ph",
    reportNumber = "FERMILAB-PUB-21-344-T, FERMILAB-PUB-21-344-T",
    doi = "10.1103/PhysRevD.105.095005",
    journal = "Phys. Rev. D",
    volume = "105",
    number = "9",
    pages = "095005",
    year = "2022"
}

@article{Buckley:2024eoe,
    author = "Buckley, Matthew R. and Fernandez, Nicolas",
    title = "{Force-feeding Supermassive Black Holes with Dissipative Dark Matter}",
    eprint = "2410.06252",
    archivePrefix = "arXiv",
    primaryClass = "hep-ph",
    month = "10",
    year = "2024"
}

@article{Chacko:2016kgg,
    author = "Chacko, Zackaria and Cui, Yanou and Hong, Sungwoo and Okui, Takemichi and Tsai, Yuhsin",
    title = "{Partially Acoustic Dark Matter, Interacting Dark Radiation, and Large Scale Structure}",
    eprint = "1609.03569",
    archivePrefix = "arXiv",
    primaryClass = "astro-ph.CO",
    doi = "10.1007/JHEP12(2016)108",
    journal = "JHEP",
    volume = "12",
    pages = "108",
    year = "2016"
}

@article{Buen-Abad:2022kgf,
    author = "Buen-Abad, Manuel A. and Chacko, Zackaria and Kilic, Can and Marques-Tavares, Gustavo and Youn, Taewook",
    title = "{Stepped partially acoustic dark matter, large scale structure, and the Hubble tension}",
    eprint = "2208.05984",
    archivePrefix = "arXiv",
    primaryClass = "hep-ph",
    reportNumber = "UTTG-10-2022",
    doi = "10.1007/JHEP06(2023)012",
    journal = "JHEP",
    volume = "06",
    pages = "012",
    year = "2023"
}

@article{Buen-Abad:2024tlb,
    author = "Buen-Abad, Manuel A. and Chacko, Zackaria and Flood, Ina and Kilic, Can and Marques-Tavares, Gustavo and Youn, Taewook",
    title = "{Atomic dark matter, interacting dark radiation, and the Hubble tension}",
    eprint = "2411.08097",
    archivePrefix = "arXiv",
    primaryClass = "hep-ph",
    reportNumber = "UTWI-37-2024",
    doi = "10.1007/JHEP07(2025)084",
    journal = "JHEP",
    volume = "07",
    pages = "084",
    year = "2025"
}

@article{Gemmell:2023trd,
    author = "Gemmell, Caleb and Roy, Sandip and Shen, Xuejian and Curtin, David and Lisanti, Mariangela and Murray, Norman and Hopkins, Philip F.",
    title = "{Dissipative Dark Substructure: The Consequences of Atomic Dark Matter on Milky Way Analog Subhalos}",
    eprint = "2311.02148",
    archivePrefix = "arXiv",
    primaryClass = "astro-ph.GA",
    doi = "10.3847/1538-4357/ad3823",
    journal = "Astrophys. J.",
    volume = "967",
    number = "1",
    pages = "21",
    year = "2024"
}

@article{Hughes:2023tcn,
    author = "Hughes, Ellie and Ge, Fei and Cyr-Racine, Francis-Yan and Knox, Lloyd and Raghunathan, Srinivasan",
    title = "{Cool dark sector, concordance, and a low {\ensuremath{\sigma}}8}",
    eprint = "2311.05678",
    archivePrefix = "arXiv",
    primaryClass = "astro-ph.CO",
    doi = "10.1103/PhysRevD.109.103516",
    journal = "Phys. Rev. D",
    volume = "109",
    number = "10",
    pages = "103516",
    year = "2024"
}

@article{Barron:2025dys,
    author = "Barron, Jared and Curtin, David and Liu, Hongwan and Munoz, Julian and Roy, Sandip",
    title = "{Constraining Dark Acoustic Oscillations with the High-Redshift UV Luminosity Function}",
    eprint = "2512.01998",
    archivePrefix = "arXiv",
    primaryClass = "astro-ph.CO",
    month = "12",
    year = "2025"
}

@article{LVK:2022ydq,
    author = "Abbott, R. and others",
    collaboration = "LVK",
    title = "{Search for subsolar-mass black hole binaries in the second part of Advanced LIGO{\textquoteright}s and Advanced Virgo{\textquoteright}s third observing run}",
    eprint = "2212.01477",
    archivePrefix = "arXiv",
    primaryClass = "astro-ph.HE",
    doi = "10.1093/mnras/stad588",
    journal = "Mon. Not. Roy. Astron. Soc.",
    volume = "524",
    number = "4",
    pages = "5984--5992",
    year = "2023",
    note = "[Erratum: Mon.Not.Roy.Astron.Soc. 526, 6234 (2023)]"
}

@article{Gurian_2022aas,
  doi = {10.3847/2041-8213/ac997c},
  url = {https://dx.doi.org/10.3847/2041-8213/ac997c},
  year = {2022},
  month = {oct},
  publisher = {The American Astronomical Society},
  volume = {939},
  number = {1},
  pages = {L12},
  author = {Gurian, James and Ryan, Michael and Schon, Sarah and Jeong, Donghui and Shandera, Sarah},
  title = {A Lower Bound on the Mass of Compact Objects from Dissipative Dark Matter},
  journal = {The Astrophysical Journal Letters}
}

@article{Ghalsasi:2017,
  author       = {Ghalsasi, Akshay and McQuinn, Matthew},
  title        = {Exploring the astrophysics of dark atoms},
  journal      = {Phys. Rev. D},
  volume       = {97},
  year         = {2018},
  doi          = {10.1103/PhysRevD.97.123018},
  eprint       = {1712.04779},
  archivePrefix= {arXiv},
  primaryClass = {astro-ph.CO}
}

@article{ACT:2025tim,
    author = "Calabrese, Erminia and others",
    collaboration = "ACT",
    title = "{The Atacama Cosmology Telescope: DR6 Constraints on Extended Cosmological Models}",
    eprint = "2503.14454",
    archivePrefix = "arXiv",
    primaryClass = "astro-ph.CO",
    reportNumber = "FERMILAB-PUB-25-0157-PPD",
    month = "3",
    year = "2025"
}

@article{ACT:2025fju,
    author = "Louis, Thibaut and others",
    collaboration = "ACT",
    title = "{The Atacama Cosmology Telescope: DR6 Power Spectra, Likelihoods and $\Lambda$ CDM Parameters}",
    eprint = "2503.14452",
    archivePrefix = "arXiv",
    primaryClass = "astro-ph.CO",
    reportNumber = "FERMILAB-PUB-25-0071-PPD",
    month = "3",
    year = "2025"
}

@article{Planck:2019nip,
    author = "Aghanim, N. and others",
    collaboration = "Planck",
    title = "{Planck 2018 results. V. CMB power spectra and likelihoods}",
    eprint = "1907.12875",
    archivePrefix = "arXiv",
    primaryClass = "astro-ph.CO",
    doi = "10.1051/0004-6361/201936386",
    journal = "Astron. Astrophys.",
    volume = "641",
    pages = "A5",
    year = "2020"
}

@article{Planck:2018vyg,
    author = "Aghanim, N. and others",
    collaboration = "Planck",
    title = "{Planck 2018 results. VI. Cosmological parameters}",
    eprint = "1807.06209",
    archivePrefix = "arXiv",
    primaryClass = "astro-ph.CO",
    doi = "10.1051/0004-6361/201833910",
    journal = "Astron. Astrophys.",
    volume = "641",
    pages = "A6",
    year = "2020",
    note = "[Erratum: Astron.Astrophys. 652, C4 (2021)]"
}

@article{Carron:2022eyg,
    author = "Carron, Julien and Mirmelstein, Mark and Lewis, Antony",
    title = "{CMB lensing from Planck PR4~maps}",
    eprint = "2206.07773",
    archivePrefix = "arXiv",
    primaryClass = "astro-ph.CO",
    doi = "10.1088/1475-7516/2022/09/039",
    journal = "JCAP",
    volume = "09",
    pages = "039",
    year = "2022"
}

@article{ACT:2023kun,
    author = "Madhavacheril, Mathew S. and others",
    collaboration = "ACT",
    title = "{The Atacama Cosmology Telescope: DR6 Gravitational Lensing Map and Cosmological Parameters}",
    eprint = "2304.05203",
    archivePrefix = "arXiv",
    primaryClass = "astro-ph.CO",
    reportNumber = "FERMILAB-PUB-23-206-PPD",
    doi = "10.3847/1538-4357/acff5f",
    journal = "Astrophys. J.",
    volume = "962",
    number = "2",
    pages = "113",
    year = "2024"
}

@article{ACT:2023dou,
    author = "Qu, Frank J. and others",
    collaboration = "ACT",
    title = "{The Atacama Cosmology Telescope: A Measurement of the DR6 CMB Lensing Power Spectrum and Its Implications for Structure Growth}",
    eprint = "2304.05202",
    archivePrefix = "arXiv",
    primaryClass = "astro-ph.CO",
    reportNumber = "FERMILAB-PUB-23-237-PPD, FERMILAB-PUB-23-237-PPD",
    doi = "10.3847/1538-4357/acfe06",
    journal = "Astrophys. J.",
    volume = "962",
    number = "2",
    pages = "112",
    year = "2024"
}

@article{Scolnic:2021amr,
    author = "Scolnic, Dan and others",
    title = "{The Pantheon+ Analysis: The Full Data Set and Light-curve Release}",
    eprint = "2112.03863",
    archivePrefix = "arXiv",
    primaryClass = "astro-ph.CO",
    doi = "10.3847/1538-4357/ac8b7a",
    journal = "Astrophys. J.",
    volume = "938",
    number = "2",
    pages = "113",
    year = "2022"
}

@article{Brout:2022vxf,
    author = "Brout, Dillon and others",
    title = "{The Pantheon+ Analysis: Cosmological Constraints}",
    eprint = "2202.04077",
    archivePrefix = "arXiv",
    primaryClass = "astro-ph.CO",
    doi = "10.3847/1538-4357/ac8e04",
    journal = "Astrophys. J.",
    volume = "938",
    number = "2",
    pages = "110",
    year = "2022"
}

@article{DESI:2024uvr,
    author = "Adame, A. G. and others",
    collaboration = "DESI",
    title = "{DESI 2024 III: baryon acoustic oscillations from galaxies and quasars}",
    eprint = "2404.03000",
    archivePrefix = "arXiv",
    primaryClass = "astro-ph.CO",
    reportNumber = "FERMILAB-PUB-24-0159-PPD",
    doi = "10.1088/1475-7516/2025/04/012",
    journal = "JCAP",
    volume = "04",
    pages = "012",
    year = "2025"
}

@article{DESI:2024lzq,
    author = "Adame, A. G. and others",
    collaboration = "DESI",
    title = "{DESI 2024 IV: Baryon Acoustic Oscillations from the Lyman alpha forest}",
    eprint = "2404.03001",
    archivePrefix = "arXiv",
    primaryClass = "astro-ph.CO",
    reportNumber = "FERMILAB-PUB-24-0147-PPD",
    doi = "10.1088/1475-7516/2025/01/124",
    journal = "JCAP",
    volume = "01",
    pages = "124",
    year = "2025"
}

@article{DESI:2024mwx,
    author = "Adame, A. G. and others",
    collaboration = "DESI",
    title = "{DESI 2024 VI: cosmological constraints from the measurements of baryon acoustic oscillations}",
    eprint = "2404.03002",
    archivePrefix = "arXiv",
    primaryClass = "astro-ph.CO",
    reportNumber = "FERMILAB-PUB-24-0154-PPD",
    doi = "10.1088/1475-7516/2025/02/021",
    journal = "JCAP",
    volume = "02",
    pages = "021",
    year = "2025"
}

@article{Lewis:2002ah,
      author         = "Lewis, Antony and Bridle, Sarah",
      title          = "{Cosmological parameters from CMB and other data: A Monte
                        Carlo approach}",
      journal        = "Phys. Rev.",
      volume         = "D66",
      year           = "2002",
      pages          = "103511",
      doi            = "10.1103/PhysRevD.66.103511",
      eprint         = "astro-ph/0205436",
      archivePrefix  = "arXiv",
      primaryClass   = "astro-ph",
      SLACcitation   = "%%CITATION = ASTRO-PH/0205436;%%",
      url            = {https://arxiv.org/abs/astro-ph/0205436}
}

@article{Lewis:2013hha,
      author         = "Lewis, Antony",
      title          = "{Efficient sampling of fast and slow cosmological
                        parameters}",
      journal        = "Phys. Rev.",
      volume         = "D87",
      year           = "2013",
      number         = "10",
      pages          = "103529",
      doi            = "10.1103/PhysRevD.87.103529",
      eprint         = "1304.4473",
      archivePrefix  = "arXiv",
      primaryClass   = "astro-ph.CO",
      SLACcitation   = "%%CITATION = ARXIV:1304.4473;%%",
      url            = {https://arxiv.org/abs/1304.4473}
}

@article{Torrado:2020dgo,
    author = "Torrado, Jesus and Lewis, Antony",
    title = "{Cobaya: Code for Bayesian Analysis of hierarchical physical models}",
    eprint = "2005.05290",
    archivePrefix = "arXiv",
    primaryClass = "astro-ph.IM",
    reportNumber = "TTK-20-15",
    doi = "10.1088/1475-7516/2021/05/057",
    journal = "JCAP",
    volume = "05",
    pages = "057",
    year = "2021"
}

@ARTICLE{CLASS_overview,
       author = {Lesgourgues, Julien},
        title = "{The Cosmic Linear Anisotropy Solving System (CLASS) I: Overview}",
      journal = {arXiv e-prints},
         year = 2011,
        month = apr,
          doi = {10.48550/arXiv.1104.2932},
archivePrefix = {arXiv},
       eprint = {1104.2932},
 primaryClass = {astro-ph.IM},
       adsurl = {https://ui.adsabs.harvard.edu/abs/2011arXiv1104.2932L}
}

@article{Blas:2011rf,
      author         = "Blas, Diego and Lesgourgues, Julien and Tram, Thomas",
      title          = "{The Cosmic Linear Anisotropy Solving System (CLASS) II:
                        Approximation schemes}",
      journal        = "JCAP",
      volume         = "1107",
      year           = "2011",
      pages          = "034",
      doi            = "10.1088/1475-7516/2011/07/034",
      eprint         = "1104.2933",
      archivePrefix  = "arXiv",
      primaryClass   = "astro-ph.CO",
      reportNumber   = "CERN-PH-TH-2011-082, LAPTH-010-11",
      SLACcitation   = "%%CITATION = ARXIV:1104.2933;%%"
}

@article{DarkKromeI,
  author       = {Michael Ryan and James Gurian and Sarah Shandera and Donghui Jeong},
  title        = {Molecular Chemistry for Dark Matter},
  journal      = {arXiv e-prints},
  volume       = {arXiv:2106.13245},
  year         = {2021},
  note         = {Submitted 24 Jun 2021, revised 5 Aug 2022},
  doi          = {10.48550/arXiv.2106.13245},
}

@article{DarkKromeII,
  author       = {James Gurian and Donghui Jeong and Michael Ryan and Sarah Shandera},
  title        = {Molecular Chemistry for Dark Matter II: Recombination, Molecule Formation, and Halo Mass Function in Atomic Dark Matter},
  journal      = {arXiv e-prints},
  volume       = {arXiv:2110.11964},
  year         = {2021},
  doi          = {10.48550/arXiv.2110.11964}
}

@article{Hou:2011ec,
    author = "Hou, Zhen and Keisler, Ryan and Knox, Lloyd and Millea, Marius and Reichardt, Christian",
    title = "{How Massless Neutrinos Affect the Cosmic Microwave Background Damping Tail}",
    eprint = "1104.2333",
    archivePrefix = "arXiv",
    primaryClass = "astro-ph.CO",
    doi = "10.1103/PhysRevD.87.083008",
    journal = "Phys. Rev. D",
    volume = "87",
    pages = "083008",
    year = "2013"
}

@article{Riess_2022,
doi = {10.3847/2041-8213/ac5c5b},
url = {https://doi.org/10.3847/2041-8213/ac5c5b},
year = {2022},
month = {jul},
publisher = {The American Astronomical Society},
volume = {934},
number = {1},
pages = {L7},
author = {Riess, Adam G. and Yuan, Wenlong and Macri, Lucas M. and Scolnic, Dan and Brout, Dillon and Casertano, Stefano and Jones, David O. and Murakami, Yukei and Anand, Gagandeep S. and Breuval, Louise and Brink, Thomas G. and Filippenko, Alexei V. and Hoffmann, Samantha and Jha, Saurabh W. and D'arcy Kenworthy, W. and Mackenty, John and Stahl, Benjamin E. and Zheng, WeiKang},
title = {A Comprehensive Measurement of the Local Value of the Hubble Constant with 1 km s-1 Mpc-1 Uncertainty from the Hubble Space Telescope and the SH0ES Team},
journal = {The Astrophysical Journal Letters}
}

@article{Kamionkowski:2022pkx,
    author = "Kamionkowski, Marc and Riess, Adam G.",
    title = "{The Hubble Tension and Early Dark Energy}",
    eprint = "2211.04492",
    archivePrefix = "arXiv",
    primaryClass = "astro-ph.CO",
    doi = "10.1146/annurev-nucl-111422-024107",
    journal = "Ann. Rev. Nucl. Part. Sci.",
    volume = "73",
    pages = "153--180",
    year = "2023"
}

@article{Finkelstein_2023,
doi = {10.3847/2041-8213/acade4},
url = {https://doi.org/10.3847/2041-8213/acade4},
year = {2023},
month = {mar},
publisher = {The American Astronomical Society},
volume = {946},
number = {1},
pages = {L13},
author = {Finkelstein, Steven L. and Bagley, Micaela B. and Ferguson, Henry C. and Wilkins, Stephen M. and Kartaltepe, Jeyhan S. and Papovich, Casey and Yung, L. Y. Aaron and Arrabal Haro, Pablo and Behroozi, Peter and Dickinson, Mark and Kocevski, Dale D. and Koekemoer, Anton M. and Larson, Rebecca L. and Le Bail, Aurélien and Morales, Alexa M. and Pérez-González, Pablo G. and Burgarella, Denis and Davé, Romeel and Hirschmann, Michaela and Somerville, Rachel S. and Wuyts, Stijn and Bromm, Volker and Casey, Caitlin M. and Fontana, Adriano and Fujimoto, Seiji and Gardner, Jonathan P. and Giavalisco, Mauro and Grazian, Andrea and Grogin, Norman A. and Hathi, Nimish P. and Hutchison, Taylor A. and Jha, Saurabh W. and Jogee, Shardha and Kewley, Lisa J. and Kirkpatrick, Allison and Long, Arianna S. and Lotz, Jennifer M. and Pentericci, Laura and Pierel, Justin D. R. and Pirzkal, Nor and Ravindranath, Swara and Ryan, Russell E. and Trump, Jonathan R. and Yang, Guang and Bhatawdekar, Rachana and Bisigello, Laura and Buat, Véronique and Calabrò, Antonello and Castellano, Marco and Cleri, Nikko J. and Cooper, M. C. and Croton, Darren and Daddi, Emanuele and Dekel, Avishai and Elbaz, David and Franco, Maximilien and Gawiser, Eric and Holwerda, Benne W. and Huertas-Company, Marc and Jaskot, Anne E. and Leung, Gene C. K. and Lucas, Ray A. and Mobasher, Bahram and Pandya, Viraj and Tacchella, Sandro and Weiner, Benjamin J. and Zavala, Jorge A.},
title = {CEERS Key Paper. I. An Early Look into the First 500 Myr of Galaxy Formation with JWST},
journal = {The Astrophysical Journal Letters}
}

@article{Harikane_2023,
doi = {10.3847/1538-4365/acaaa9},
url = {https://doi.org/10.3847/1538-4365/acaaa9},
year = {2023},
month = {feb},
publisher = {The American Astronomical Society},
volume = {265},
number = {1},
pages = {5},
author = {Harikane, Yuichi and Ouchi, Masami and Oguri, Masamune and Ono, Yoshiaki and Nakajima, Kimihiko and Isobe, Yuki and Umeda, Hiroya and Mawatari, Ken and Zhang, Yechi},
title = {A Comprehensive Study of Galaxies at z ~ 9–16 Found in the Early JWST Data: Ultraviolet Luminosity Functions and Cosmic Star Formation History at the Pre-reionization Epoch},
journal = {The Astrophysical Journal Supplement Series}
}

@article{bouwens_evolution_2023,
	title = {Evolution of the {UV} {LF} from z ~ 15 to z ~ 8 using new {JWST} {NIRCam} medium-band observations over the {HUDF}/{XDF}},
	volume = {523},
	issn = {0035-8711},
	url = {https://doi.org/10.1093/mnras/stad1145},
	doi = {10.1093/mnras/stad1145},
	number = {1},
	journal = {Monthly Notices of the Royal Astronomical Society},
	author = {Bouwens, Rychard J and Stefanon, Mauro and Brammer, Gabriel and Oesch, Pascal A and Herard-Demanche, Thomas and Illingworth, Garth D and Matthee, Jorryt and Naidu, Rohan P and van Dokkum, Pieter G and van Leeuwen, Ivana F},
	month = apr,
	year = {2023},
	note = {\_eprint: https://academic.oup.com/mnras/article-pdf/523/1/1036/50488670/stad1145.pdf},
	pages = {1036--1055},
}

@article{adame_desi_2025,
	title = {{DESI} 2024 {VI}: cosmological constraints from the measurements of baryon acoustic oscillations},
	volume = {2025},
	url = {https://doi.org/10.1088/1475-7516/2025/02/021},
	doi = {10.1088/1475-7516/2025/02/021},
	number = {02},
	journal = {Journal of Cosmology and Astroparticle Physics},
	author = {Adame, A.G. and Aguilar, J. and Ahlen, S. and Alam, S. and Alexander, D.M. and Alvarez, M. and Alves, O. and Anand, A. and Andrade, U. and Armengaud, E. and Avila, S. and Aviles, A. and Awan, H. and Bahr-Kalus, B. and Bailey, S. and Baltay, C. and Bault, A. and Behera, J. and BenZvi, S. and Bera, A. and Beutler, F. and Bianchi, D. and Blake, C. and Blum, R. and Brieden, S. and Brodzeller, A. and Brooks, D. and Buckley-Geer, E. and Burtin, E. and Calderon, R. and Canning, R. and Carnero Rosell, A. and Cereskaite, R. and Cervantes-Cota, J.L. and Chabanier, S. and Chaussidon, E. and Chaves-Montero, J. and Chen, S. and Chen, X. and Claybaugh, T. and Cole, S. and Cuceu, A. and Davis, T.M. and Dawson, K. and de la Macorra, A. and de Mattia, A. and Deiosso, N. and Dey, A. and Dey, B. and Ding, Z. and Doel, P. and Edelstein, J. and Eftekharzadeh, S. and Eisenstein, D.J. and Elliott, A. and Fagrelius, P. and Fanning, K. and Ferraro, S. and Ereza, J. and Findlay, N. and Flaugher, B. and Font-Ribera, A. and Forero-Sánchez, D. and Forero-Romero, J.E. and Frenk, C.S. and Garcia-Quintero, C. and Gaztañaga, E. and Gil-Marín, H. and Gontcho, S.Gontcho A. and Gonzalez-Morales, A.X. and Gonzalez-Perez, V. and Gordon, C. and Green, D. and Gruen, D. and Gsponer, R. and Gutierrez, G. and Guy, J. and Hadzhiyska, B. and Hahn, C. and Hanif, M.M.S. and Herrera-Alcantar, H.K. and Honscheid, K. and Howlett, C. and Huterer, D. and Iršič, V. and Ishak, M. and Juneau, S. and Karaçaylı, N.G. and Kehoe, R. and Kent, S. and Kirkby, D. and Kremin, A. and Krolewski, A. and Lai, Y. and Lan, T.-W. and Landriau, M. and Lang, D. and Lasker, J. and Le Goff, J.M. and Le Guillou, L. and Leauthaud, A. and Levi, M.E. and Li, T.S. and Linder, E. and Lodha, K. and Magneville, C. and Manera, M. and Margala, D. and Martini, P. and Maus, M. and McDonald, P. and Medina-Varela, L. and Meisner, A. and Mena-Fernández, J. and Miquel, R. and Moon, J. and Moore, S. and Moustakas, J. and Mueller, E. and Muñoz-Gutiérrez, A. and Myers, A.D. and Nadathur, S. and Napolitano, L. and Neveux, R. and Newman, J.A. and Nguyen, N.M. and Nie, J. and Niz, G. and Noriega, H.E. and Padmanabhan, N. and Paillas, E. and Palanque-Delabrouille, N. and Pan, J. and Penmetsa, S. and Percival, W.J. and Pieri, M.M. and Pinon, M. and Poppett, C. and Porredon, A. and Prada, F. and Pérez-Fernández, A. and Pérez-Ràfols, I. and Rabinowitz, D. and Raichoor, A. and Ramírez-Pérez, C. and Ramirez-Solano, S. and Rashkovetskyi, M. and Ravoux, C. and Rezaie, M. and Rich, J. and Rocher, A. and Rockosi, C. and Roe, N.A. and Rosado-Marin, A. and Ross, A.J. and Rossi, G. and Ruggeri, R. and Ruhlmann-Kleider, V. and Samushia, L. and Sanchez, E. and Saulder, C. and Schlafly, E.F. and Schlegel, D. and Schubnell, M. and Seo, H. and Shafieloo, A. and Sharples, R. and Silber, J. and Slosar, A. and Smith, A. and Sprayberry, D. and Tan, T. and Tarlé, G. and Taylor, P. and Trusov, S. and Ureña-López, L.A. and Vaisakh, R. and Valcin, D. and Valdes, F. and Vargas-Magaña, M. and Verde, L. and Walther, M. and Wang, B. and Wang, M.S. and Weaver, B.A. and Weaverdyck, N. and Wechsler, R.H. and Weinberg, D.H. and White, M. and Yu, J. and Yu, Y. and Yuan, S. and Yèche, C. and Zaborowski, E.A. and Zarrouk, P. and Zhang, H. and Zhao, C. and Zhao, R. and Zhou, R. and Zhuang, T. and Zou, H. and collaboration, The DESI},
	month = feb,
	year = {2025},
	note = {Publisher: IOP Publishing},
	pages = {021},
}

@article{w4c6-1r5j,
  title = {Extended dark energy analysis using DESI DR2 BAO measurements},
  author = {Lodha, K. and Calderon, R. and Matthewson, W. L. and Shafieloo, A. and Ishak, M. and Pan, J. and Garcia-Quintero, C. and Huterer, D. and Valogiannis, G. and Ure\~na-L\'opez, L. A. and Kamble, N. V. and Parkinson, D. and Kim, A. G. and Zhao, G. B. and Cervantes-Cota, J. L. and Rohlf, J. and Lozano-Rodr\'{\i}guez, F. and Rom\'an-Herrera, J. O. and Abdul-Karim, M. and Aguilar, J. and Ahlen, S. and Alves, O. and Andrade, U. and Armengaud, E. and Aviles, A. and Behera, J. and BenZvi, S. and Bianchi, D. and Brodzeller, A. and Brooks, D. and Burtin, E. and Canning, R. and Rosell, A. Carnero and Casas, L. and Castander, F. J. and Charles, M. and Chaussidon, E. and Chaves-Montero, J. and Chebat, D. and Claybaugh, T. and Cole, S. and Cuceu, A. and Dawson, K. S. and de la Macorra, A. and de Mattia, A. and Deiosso, N. and Demina, R. and Dey, Arjun and Dey, Biprateep and Ding, Z. and Doel, P. and Eisenstein, D. J. and Elbers, W. and Ferraro, S. and Font-Ribera, A. and Forero-Romero, J. E. and Garrison, Lehman H. and Gazta\~naga, E. and Gil-Mar\'{\i}n, H. and Gontcho, S. Gontcho A. and Gonzalez-Morales, A. X. and Gutierrez, G. and Guy, J. and Hahn, C. and Herbold, M. and Herrera-Alcantar, H. K. and Honscheid, K. and Howlett, C. and Juneau, S. and Kehoe, R. and Kirkby, D. and Kisner, T. and Kremin, A. and Lahav, O. and Lamman, C. and Landriau, M. and Le Guillou, L. and Leauthaud, A. and Levi, M. E. and Li, Q. and Magneville, C. and Manera, M. and Martini, P. and Meisner, A. and Mena-Fern\'andez, J. and Miquel, R. and Moustakas, J. and Santos, D. Mu\~noz and Mu\~noz-Guti\'errez, A. and Myers, A. D. and Nadathur, S. and Niz, G. and Noriega, H. E. and Paillas, E. and Palanque-Delabrouille, N. and Percival, W. J. and Pieri, Matthew M. and Poppett, C. and Prada, F. and P\'erez-Fern\'andez, A. and P\'erez-R\`afols, I. and Ram\'{\i}rez-P\'erez, C. and Rashkovetskyi, M. and Ravoux, C. and Ross, A. J. and Rossi, G. and Ruhlmann-Kleider, V. and Samushia, L. and Sanchez, E. and Schlegel, D. and Schubnell, M. and Seo, H. and Sinigaglia, F. and Sprayberry, D. and Tan, T. and Tarl\'e, G. and Taylor, P. and Turner, W. and Vargas-Maga\~na, M. and Walther, M. and Weaver, B. A. and Wolfson, M. and Y\`eche, C. and Zarrouk, P. and Zhou, R. and Zou, H.},
  collaboration = {DESI Collaboration},
  journal = {Phys. Rev. D},
  volume = {112},
  issue = {8},
  pages = {083511},
  numpages = {27},
  year = {2025},
  month = {Oct},
  publisher = {American Physical Society},
  doi = {10.1103/w4c6-1r5j},
  url = {https://link.aps.org/doi/10.1103/w4c6-1r5j}
}

@article{Bullock:2017xww,
    author = "Bullock, James S. and Boylan-Kolchin, Michael",
    title = "{Small-Scale Challenges to the $\Lambda$CDM Paradigm}",
    eprint = "1707.04256",
    archivePrefix = "arXiv",
    primaryClass = "astro-ph.CO",
    doi = "10.1146/annurev-astro-091916-055313",
    journal = "Ann. Rev. Astron. Astrophys.",
    volume = "55",
    pages = "343--387",
    year = "2017"
}

@article{Abdalla:2022yfr,
    author = "Abdalla, Elcio and others",
    title = "{Cosmology intertwined: A review of the particle physics, astrophysics, and cosmology associated with the cosmological tensions and anomalies}",
    eprint = "2203.06142",
    archivePrefix = "arXiv",
    primaryClass = "astro-ph.CO",
    reportNumber = "FERMILAB-CONF-22-192-SCD",
    doi = "10.1016/j.jheap.2022.04.002",
    journal = "JHEAp",
    volume = "34",
    pages = "49--211",
    year = "2022"
}

@book{osti_6676289,
  author       = {Peterkop, R K},
  title        = {Theory of ionization of atoms by electron impact},
  url          = {https://www.osti.gov/biblio/6676289},
  place        = {United States},
  publisher    = {Colorado Associated University Press,Boulder, CO},
  year         = {1977},
  month        = {01}}

@article{Chang:2018bgx,
    author = "Chang, Jae Hyeok and Egana-Ugrinovic, Daniel and Essig, Rouven and Kouvaris, Chris",
    title = "{Structure Formation and Exotic Compact Objects in a Dissipative Dark Sector}",
    eprint = "1812.07000",
    archivePrefix = "arXiv",
    primaryClass = "hep-ph",
    doi = "10.1088/1475-7516/2019/03/036",
    journal = "JCAP",
    volume = "03",
    pages = "036",
    year = "2019"
}

@article{Mohapatra:2001sx,
    author = "Mohapatra, R. N. and Nussinov, S. and Teplitz, V. L.",
    title = "{Mirror matter as selfinteracting dark matter}",
    eprint = "hep-ph/0111381",
    archivePrefix = "arXiv",
    reportNumber = "UMD-PP-02-023",
    doi = "10.1103/PhysRevD.66.063002",
    journal = "Phys. Rev. D",
    volume = "66",
    pages = "063002",
    year = "2002"
}

@article{Foot:2018qpw,
    author = "Foot, R.",
    title = "{Dissipative dark matter halos: The steady state solution II}",
    eprint = "1801.09359",
    archivePrefix = "arXiv",
    primaryClass = "astro-ph.GA",
    doi = "10.1103/PhysRevD.97.103006",
    journal = "Phys. Rev. D",
    volume = "97",
    number = "10",
    pages = "103006",
    year = "2018"
}

@article{Foot:2014mia,
    author = "Foot, R.",
    title = "{Mirror dark matter: Cosmology, galaxy structure and direct detection}",
    eprint = "1401.3965",
    archivePrefix = "arXiv",
    primaryClass = "astro-ph.CO",
    doi = "10.1142/S0217751X14300130",
    journal = "Int. J. Mod. Phys. A",
    volume = "29",
    pages = "1430013",
    year = "2014"
}

@article{Buckley:2017ttd,
    author = "Buckley, Matthew R. and DiFranzo, Anthony",
    title = "{Collapsed Dark Matter Structures}",
    eprint = "1707.03829",
    archivePrefix = "arXiv",
    primaryClass = "hep-ph",
    doi = "10.1103/PhysRevLett.120.051102",
    journal = "Phys. Rev. Lett.",
    volume = "120",
    number = "5",
    pages = "051102",
    year = "2018"
}

@article{Ryan:2021dis,
    author = "Ryan, Michael and Gurian, James and Shandera, Sarah and Jeong, Donghui",
    title = "{Molecular Chemistry for Dark Matter}",
    eprint = "2106.13245",
    archivePrefix = "arXiv",
    primaryClass = "astro-ph.CO",
    doi = "10.3847/1538-4357/ac75ef",
    journal = "Astrophys. J.",
    volume = "934",
    pages = "120",
    year = "2022"
}

@article{Gurian:2021qhk,
    author = "Gurian, James and Jeong, Donghui and Ryan, Michael and Shandera, Sarah",
    title = "{Molecular Chemistry for Dark Matter II: Recombination, Molecule Formation, and Halo Mass Function in Atomic Dark Matter}",
    eprint = "2110.11964",
    archivePrefix = "arXiv",
    primaryClass = "astro-ph.CO",
    doi = "10.3847/1538-4357/ac75e4",
    journal = "Astrophys. J.",
    volume = "934",
    pages = "121",
    year = "2022"
}

@article{Ryan:2021tgw,
    author = "Ryan, Michael and Shandera, Sarah and Gurian, James and Jeong, Donghui",
    title = "{Molecular Chemistry for Dark Matter III: DarkKROME}",
    eprint = "2110.11971",
    archivePrefix = "arXiv",
    primaryClass = "astro-ph.CO",
    doi = "10.3847/1538-4357/ac75e5",
    journal = "Astrophys. J.",
    volume = "934",
    pages = "122",
    year = "2022"
}

\end{document}